\documentclass[journal]{IEEEtran}
\IEEEoverridecommandlockouts                              
\usepackage{graphicx} 
\graphicspath{{figures/}}
\usepackage{amsmath} 
\usepackage{amssymb}
\usepackage{amsthm}
\usepackage{lettrine}
\usepackage{epstopdf}
\usepackage{xcolor}
\usepackage{gensymb}
\usepackage{pgfplots} 
\pgfplotsset{compat=newest} 
\pgfplotsset{plot coordinates/math parser=false} 
\usepackage[absolute,overlay]{textpos}
\usepackage{mathtools}
\usepackage{algorithm}
\usepackage{algpseudocode}
\usepackage{subfig}
\usepackage{verbatim}
\usepackage{multirow}
\usepackage{float}
\usepackage{enumerate}
\usepackage{soul}
\usepackage[font=small]{caption}
\usepackage{amsfonts}
\usepackage{cancel} 
\usepackage{siunitx}
\usepackage[hidelinks]{hyperref}
\usepackage{marginnote}
\mathtoolsset{showonlyrefs} 
\usepackage{ctable}
\noeqref{www,wwww,eq:KKT_PEVs_stat,eq:inf_inner_a,eq:inf_inner_b,eq:inf_inner_c,eq:F_N,eq,eq:apa_inner_a,eq:conclusion_proof_APA}
\noeqref{eq:ex_1_system_1,eq:ex_1_system_2,eq:ex_1_system_3,eq:ex_1_system_4,eq:ex_1_system_5,eq:ex_1_system_6}
\noeqref{eq:ex_1_system_compact_1,eq:ex_1_system_compact_2,eq:ex_1_system_compact_3,eq:ex_1_system_compact_4,eq:ex_1_system_compact_5,eq:ex_1_system_compact_6}
\noeqref{eq:KKT_VI_a,eq:KKT_VI_b,eq:KKT_VI_c}

\definecolor{Granata}{rgb}{0.64,0,0} 			%
\definecolor{QuasiBlue}{rgb}{0.03,0.3,0.72} 	%
\definecolor{PortlandGreen}{RGB}{99,166,63} 	%
\definecolor{ReadableGreen}{RGB}{0,120,0}
\definecolor{OrangeRed}{RGB}{255,69,0} 			%
\definecolor{QuasiBlue}{rgb}{0.03,0.3,0.72}
\definecolor{YellowIntense}{RGB}{240,198,31} 	%

\newcommand{\defeq}{\coloneqq}
\newcommand{\eqdef}{\eqqcolon}

\newtheorem{definition}{Definition}
\newtheorem{theorem}{Theorem}

\newtheorem{lemma}{Lemma}
\newtheorem{example}{Example}
\newtheorem{remark}{Remark}
\newtheorem{proposition}{Proposition}

\newtheorem*{sassumption}{Standing assumption}

\DeclareSymbolFont{bbold}{U}{bbold}{m}{n}
\DeclareSymbolFontAlphabet{\mathbbold}{bbold}

\newcommand{\vect}[1]{\mathbbold{#1}}
\newcommand{\zeros}[1][]{\vect{0}_{#1}}
\newcommand{\ones}[1][]{\vect{1}_{#1}}

\newcommand\R{\mathbb{R}}
\newcommand\Rn{\R^n}

\newcommand\mc[1]{\mathcal{#1}}
\newcommand\mb[1]{\mathbb{#1}}
\renewcommand\i{^i}

\newcommand{\minn}[1]{\underset{#1}{\operatorname{min}}\,}
\newcommand{\maxx}[1]{\underset{#1}{\operatorname{max}}\,}
\newcommand{\proj}[1]{\underset{#1}{\operatorname{Proj}}\,}

\newcommand{\X}{\mc{X}}
\newcommand{\N}{M}

\newcommand{\bx}{\bar x}
\newcommand{\blambda}{\bar \lambda}
\newcommand{\bmu}{\bar \mu}

\newcommand{\Lmax}{L} 
\renewcommand{\a}{\alpha}
\newcommand{\sa}{\s^\a}
\newcommand{\suma}{\sum_{\a=1}^A}
\newcommand{\RnA}{\mathbb{R}^{n A}}
\newcommand{\sj}{\s_i}
\newcommand{\sja}{\s_i^\alpha}

\newcommand{\bs}{\bar{\s}}
\newcommand{\bsj}{\bar{\s}_i}

\newcommand{\jh}{c_{ij}}

\newcommand{\jn}{_{i=1}^n}

\renewcommand{\S}{\mathcal{S}}
\newcommand{\Spl}{\S_{+}}
\newcommand{\B}{\mathcal{B}}

\newcommand{\s}{x}

\newcommand{\jj}{{i^\star}}
\newcommand{\hh}{{j^\star}}

\newcommand{\A}{\mc{A}}

\newcommand{\epss}{{\tilde{\varepsilon}}}
\renewcommand{\ss}{{\tilde{\s}}}
\newcommand{\sss}{\s^\star}

\newcommand{\cmin}{c_\tmin}
\newcommand{\muu}{\mu^\star}
\newcommand{\etaa}{\eta^\star}
\newcommand\mydots{\hbox to 1em{.\hss.\hss.}}
\newcommand{\onen}{\{1,\mydots,n\}}
\newcommand{\oneN}{\{1,\mydots,\N\}}

\newcommand{\onenjj}{\{1,\mydots,n\} \backslash \{ \jj \}}

\newcommand{\oneA}{\{1,\mydots,A\}}
\newcommand{\Rpl}{\R_{\ge 0}}
\newcommand{\Niout}{\onen}
\newcommand{\Niin}{\onen}
\newcommand{\Nenvy}{\mathcal{E}^\textup{out}}
\newcommand{\Nenvyrev}{\mathcal{E}^\textup{in}}
\newcommand{\Nienvy}{\mathcal{E}_i^\textup{out}}
\newcommand{\Nienvyrev}{\mathcal{E}_j^\textup{in}}
\newcommand{\Njenvy}{\mathcal{E}_j^\textup{out}}

\newcommand{\tmin}{\textup{min}}
\newcommand{\tr}{^\top}
\renewcommand{\ni}{\noindent}
\newcommand{\argmaxx}[1]{\underset{#1}{\operatorname{argmax}}\,}
\newcommand{\xijk}{x_{i\to j}(k)}

\makeatletter
\newcommand\footnoteref[1]{\protected@xdef\@thefnmark{\ref{#1}}\@footnotemark}
\makeatother

\newcommand*{\colorboxed}{}
\def\colorboxed#1#{%
  \colorboxedAux{#1}%
}
\newcommand*{\colorboxedAux}[3]{%
  \begingroup
    \colorlet{cb@saved}{.}%
    \color#1{#2}%
    \boxed{%
      \color{cb@saved}%
      #3%
    }%
  \endgroup
}

\title{\LARGE \bf
The Nash Equilibrium with Inertia in Population Games
}

\author{Basilio Gentile, Dario Paccagnan, Bolutife Ogunsula, and John Lygeros %
\thanks{
This work was supported by the European Commission project DYMASOS (FP7-ICT 611281), and by the SNSF Grant \#P2EZP2-181618.
}}

\begin{document}

\maketitle
\thispagestyle{plain}

\begin{abstract}
In the traditional game-theoretic set up, where agents select actions and experience corresponding utilities, an equilibrium is a configuration where no agent can improve their utility by unilaterally switching to a different action. In this work, we introduce the novel notion of inertial Nash equilibrium to account for the fact that, in many practical situations, action changes do not come for free.
 Specifically, we consider a population game and introduce the coefficients $c_{ij}$ describing the cost an agent incurs by switching from action $i$ to action $j$. We define an inertial Nash equilibrium as a distribution over the action space where no agent benefits in moving to a different action, while taking into account the cost of this change.
First, we show that the set of inertial Nash equilibria contains all the Nash equilibria, but is in general not convex.
Second, we argue that classical algorithms for computing Nash equilibria cannot be used in the presence of switching costs. We then propose a natural better-response dynamics and prove its convergence to an inertial Nash equilibrium.
We apply our results to predict the drivers' distribution of an \mbox{on-demand ride-hailing platform.}
\end{abstract}

\vspace*{-5mm}
\section{Introduction}
\label{sec:intro}
Game theory has originated as a set of tools to model and describe the interaction of multiple decision makers, or agents. The goal is typically to determine whether decision makers will come to some form of equilibrium, the most common of which is the Nash equilibrium. Informally, a set of strategies constitutes a Nash equilibrium if no agent benefits by unilaterally deviating form the current action, while the other agents stay put. This notion of equilibrium has found countless applications, among others to energy systems~\cite{ma:callaway:hiskens:13}, transmission networks~\cite{alpcan2002cdma}, commodity markets~\cite{johari2004efficiency}, traffic flow \cite{wardrop1952road}, and mechanism design~\cite{roughgarden_AGT_book}.

While the original definition of Nash equilibrium does not account for the cost incurred by agents when moving to a different action, in practical situations decision makers often incur a physical, psychological, or monetary cost for such deviation. This is the case, for example, when relocating to a new neighbourhood~\cite{clark1996households}, or when switching financial strategy in the stock market~\cite{schulmeister2008general}.
When the decision makers are humans, the psychological resistance to change has been well documented and studied at the professional and organizational level~\cite{coch1948overcoming} as well as at the individual and private level~\cite{oreg2003resistance}, or at the customer level~\cite{oyeniyi2010switching}.

To take into account such phenomena, we introduce the novel concept of \textit{inertial Nash equilibrium}.
Specifically, we consider a setup where a large number of agents choose among $n$ common actions. Agents selecting a given action receive a utility that depends only on the agents' distribution over the action space, in the same spirit of population games \cite{sandholm2001potential}.
In this context, a Nash equilibrium consists in an agent distribution over the action space for which every utilized action yields maximum utility.
The same concept was proposed in the seminal work of Wardrop for a route-choice game in road traffic networks~\cite{wardrop1952road}.
We extend this framework and model the cost incurred by any agent when moving from action $i$ to action $j$ with the non-negative coefficients $c_{ij}$. We define an \emph{inertial Nash equilibrium} as a distribution over the action space where no agent has any incentive to unilaterally change action, where the quality of an alternative action is measured by its \emph{net utility}, i.e., the corresponding utility \emph{minus} the cost of the action change.

We show that introducing such costs leads to a larger set of equilibria that is in general not convex, even if the set of Nash equilibria without switching costs is so. We argue that classical algorithms to compute a Nash equilibrium are not  suitable for computing inertial Nash equilibria, because i) they may not terminate even if already at an inertial Nash equilibrium, and ii) their execution is not compatible with the agents' rationality assumption,
as agents might be required to perform a detrimental move.
To overcome these issues, we propose an algorithm based on better-response dynamics, where agents switch action only if it is to their advantage when factoring the \mbox{cost of such change.}

\vspace*{2mm}
\ni {\bf Contributions.}
Our main contributions are as follows.
\begin{enumerate}[i)]
\item We introduce the notion of inertial Nash equilibrium (Definition~\ref{def:Nash_inertia}) and position it in the context of the existing literature, notably in relation to population games \cite{sandholm2001potential} and more specifically migration equilibria \cite{nagurney1989migration}.
\item We show that the set of inertial Nash equilibria can be equivalently characterized through a variational inequality (Theorem~\ref{thm:equiv_VI}) and we prove a strong negative result: the operator that arises in the resulting variational inequality is non-monotone in all  the meaningful instances of the inertial Nash equilibrium problem (Theorem~\ref{thm:not_MON}). This implies that existing algorithms for computing equilibria based on the solution of variational inequalities are in general not suitable for computing an inertial Nash equilibrium.
\item We propose and analyse a novel algorithm and prove its convergence to an inertial Nash equilibrium under weak assumptions (Theorem~\ref{thm:convergence_alg_distr}).
\end{enumerate}

\vspace*{2mm}
\ni {\bf Organization.} In Section~\ref{sec:inertial_Nash} we introduce the notion of inertial Nash equilibrium, and show its non-uniqueness as well as the non-convexity of the equilibrium set. A comparison with related works is presented in Section~\ref{sec:literature}. In Section~\ref{sec:VI_reformulation} we reformulate the inertial Nash equilibrium problem as a variational inequality, study the monotonicity properties of the corresponding operator (more precisely, the lack thereof), and present the issues associated with the use of existing algorithms. In Section~\ref{sec:algorithm_intertial} we propose a modified best-response dynamics that provably converges to an inertial Nash equilibrium.
Extensions of the model are presented in Section~\ref{sec:extensions}.
In Section~\ref{sec:application} we validate our model with a numerical study of area coverage for on demand ride-hailing in Hong Kong.
Appendix~\ref{app:VI} provides background material, while all the proofs are reported in Appendix~\ref{app:proofs}.

\vspace*{2mm}
\ni {\bf Notation.}
The space of $n$-dimensional real vectors is denoted with $\Rn$, while $\Rpl^n$ is the space of non-negative $n$-dimensional real vectors and $\R_{>0}^n$ is the space of strictly positive $n$-dimensional real vectors.
The symbol $\ones[n]$ indicates the $n$-dimensional vector of unit entries, whereas $\zeros[n]$ is the $n$-dimensional vector of zero entries.
If $x,y\in\Rn$, the notation $x \ge y$ indicates that $x_j\ge y_j$ for all $j\in\onen$. The vector $\mathbf{e}_i$ denotes the $i^\text{th}$ vector of the canonical basis.
Given $A\in\mathbb{R}^{n\times n}$, $A\succ0$ ($\succeq0$) if and only if $x^\top A x = \frac{1}{2}x\tr(A+A\tr)x >0~(\ge0),$ for all  $x\neq \zeros[n]$.
$\textup{blkdiag}(A_1,\dots,A_\N)$ is the block diagonal matrix with blocks $A_1, \dots, A_\N$. $\|A\|$ is the induced $2$-norm on $A$. Given $g(x):\mathbb{R}^n \rightarrow \mathbb{R}^m$ we define $\nabla_x g(x) \in \mathbb{R}^{n\times m}$ with $[\nabla_x g(x)]_{i,j}\coloneqq \frac{\partial g_j(x)}{\partial x\i}$.
If $n=m=1$, we use $g'(x)$ to denote the derivative of $g$ at the point $x$. $I_n$ denotes the $n\times n$ identity matrix. $\proj{\mathcal{X}}[x]$ is the Euclidean projection of the vector $x$ onto a closed and convex set $\mathcal{X}$.

\section{Inertial Nash equilibrium: definition and examples}
\label{sec:inertial_Nash}
\subsection{Definition of Inertial Equilibrium}
We consider a large number of competing agents with a finite set of common actions $\{1,\dots,n\}$. For selecting action $i\in \{1,\dots,n\}$, an agent receives a utility $u_i(x)$, where $x=[x_1,\dots,x_n]$, and $x_i$ denotes the fraction of agents selecting action $i$.  Observe that, with the introduction of the utility functions $u_i:\mb{R}^n_{\ge0}\rightarrow\mb{R}$, we are implicitly assuming that the utility received by playing action $i$ only depends on the distribution of the agents, and not on which agent selected which action, a modelling assumption typically employed in population games \cite{sandholm2001potential}. Within this framework, a Nash equilibrium is a distribution over the action space where no agent has any incentive in deviating to a different action. This requirement can be formalized by introducing the unit simplex\footnote{The formulation with unitary mass and the corresponding results seamlessly generalise to agents of combined mass $\gamma>0$.} in dimension $n$, denoted with $\S$, and its relative interior $\Spl$
\[
\begin{split}
\S &\coloneqq\{\s \in\mb{R}^n~\text{s.t.}~\s\ge0,~\ones[n]^\top \s=1\},\\
\Spl &\coloneqq\{\s \in\mb{R}^n~\text{s.t.}~\s>0,~\ones[n]^\top \s=1\}.	
\end{split}
\]

\begin{definition}[Nash equilibrium, \cite{wardrop1952road}]
\label{def:Nash_parallel}
Given $n$ utilities $\{u_i\}_{i=1}^n$ with $u_i:\Rpl^n\to\R$, the vector $\bs \in \S$ is a \mbox{\textit{Nash equilibrium} if}
\begin{equation}
\label{eq:no_improv_parallel}
\bsj > 0 \implies u_i(\bs) \ge u_j(\bs), \quad \forall \, i,\,j \in\onen.
\end{equation}
\end{definition}
\ni Despite being widely used in the applications, Definition \ref{def:Nash_parallel} does not account for the cost associated with an action switch.
We extend the previous model by introducing the non-negative coefficients $c_{ij}$ to represent the cost experienced by any agent when moving from action $i$ to $j$.
We then define an inertial Nash equilibrium as a distribution over the action space where no agent can benefit by moving to a different action, while taking into account the cost of such change.

\begin{definition}[Inertial Nash equilibrium]
\label{def:Nash_inertia}
Given $n$ utilities $\{u_i\}_{i=1}^n$, $u_i:\Rpl^n\to\R$, $n^2$ non-negative switching costs $\{\jh\}_{i,j=1}^n$, the vector $\bs \in \S$ is an \textit{Inertial Nash equilibrium} if
\begin{equation}
\bsj > 0 \implies u_i(\bs) \ge u_j(\bs)-c_{ij}, \quad \forall \, i,\,j \in\onen.
\label{eq:no_improv}
\end{equation}
\end{definition}

\ni
In the remainder of this manuscript we focus on problems where there is no cost for staying put, as formalized next.
\begin{sassumption}
The switching costs satisfy $c_{ii}=0$ for all $i\in\{1,\dots,n\}$.
\end{sassumption}

Observe that conditions \eqref{eq:no_improv_parallel} and \eqref{eq:no_improv} do not impose any constraint on actions that are not currently selected by any agent (i.e., those with $\bs_i=0$). In other words, the utility of one such action can be arbitrarily low, and the configuration $\bs$ still be an equilibrium.
Despite being a natural extension to the traditional notions of equilibrium in game theory, to the best of our knowledge, Definition~\ref{def:Nash_inertia} is novel. Its relevance stems from the observation that the coefficients $c_{ij}$ can model different and common phenomena, such as:
\begin{itemize}
\item[-] the tendency of agents to adhere to their habits, or their reluctance to try something different;
\item[-] actual costs or fees that agents incur for switching action;
\item[-] the lack of accurate information about other options.
\end{itemize}

In the following, we provide two examples of problems that can be captured within this framework.\\[2mm]
{\bf On demand ride-hailing.} Ride-hailing systems are platforms that allow customers to travel from a given origin to a desired destination, typically within the same city. Examples include taxi companies as well as platforms such as Uber, Lyft or Didi. In our framework, the drivers correspond to agents and geographical locations to available actions. Each utility describes the profitability of a given location, which depends on the arrival rate of customers in that location, and on the fraction of vehicles available in that same location. The cost (fuel and time) that a driver incurs while moving between two different physical locations is captured by $c_{ij}$. Such model can predict how drivers distribute themselves over the city.\\[2mm]
{\bf Task assignment in server network.} We are given a finite number of geographically dispersed servers represented with nodes, and connected through a network. Each server corresponds to an action $i\in\{1,\dots,n\}$. A large number of agents has a list of jobs that originates in various nodes on the network and wishes to execute this list as swiftly as possible. The speed at which each server can process a job depends on the load on the server and is captured by $u_i(x_i)$. Moving a job between server $i$ and $j$ requires an amount of time and resources captured by $c_{ij}$. This model can predict how agents distribute their jobs over the set of servers.

\vspace*{2mm}
We note that the set of inertial Nash equilibria contains the set of Nash equilibria, due to the non negativity of $c_{ij}$.

\begin{lemma}
\label{lem:parallel_is_inertial}
Every Nash equilibrium is an inertial Nash equilibrium.
\end{lemma}
The proof follows from Definition \ref{def:Nash_parallel} and \ref{def:Nash_inertia}, since condition \eqref{eq:no_improv_parallel} implies condition \eqref{eq:no_improv}, as $c_{ij}\ge0$ for all $i,j\in\{1,\dots,n\}$.
In the following we refer to an (inertial) Nash equilibrium as just an (inertial) equilibrium.
\subsection{Non-uniqueness and non-convexity of the equilibrium set}
The following example shows that the set of inertial equilibria is in general neither convex, nor a singleton. This will pose significant algorithmic challenges, as discussed in Section \ref{sec:VI_reformulation}.

\begin{example}
\label{ex:inertial_1}
Let $n=3$, and consider utilities and switching costs of the form%
\begin{align}
&u_1(\s) = 1.2-\s_1 \\[-0.5cm]
&u_2(\s) = 1.2-\s_2 & & C = \begin{bmatrix} 0 & 0.2 & 0.3 \\ 1 & 0 & 0.8 \\ 0.1 & 1.2 & 0 \end{bmatrix},
\label{eq:ex_non_convex_1_utilities}
\\[-0.5cm]
&u_3(\s) = 1-\s_3,
\end{align}
where the entry $(i,j)$ of $C$ equals $\jh$. Note that $\s_3=1-\s_1-\s_2$. The equilibrium conditions~\eqref{eq:no_improv}  then become
{\begin{subequations}
\begin{align}
&\s_1 > 0 & & \Rightarrow 	& & \colorboxed{YellowIntense}{\s_2 \ge \s_1 - 0.2}	\label{eq:ex_1_system_1} \\
&\s_1 > 0 & & \Rightarrow 	& & \colorboxed{PortlandGreen}{\s_2 \le -2\s_1 + 1.5}   \label{eq:ex_1_system_2} \\
&\s_2 > 0 & & \Rightarrow 	& & {\s_2 \le \s_1+1}			\label{eq:ex_1_system_3} \\
&\s_2 > 0 & & \Rightarrow 	& & {\s_2 \le -0.5 \s_1 + 1}   \label{eq:ex_1_system_4} \\
&\s_3 > 0 & & \Rightarrow 	& & \colorboxed{red}{\s_2 \ge -2\s_1+1.1}   \label{eq:ex_1_system_5} \\
&\s_3 > 0 & & \Rightarrow 	& & {2\s_2 \ge -\s_1},
\label{eq:ex_1_system_6}
\end{align}
\label{eq:ex_1_system}
\end{subequations}
}

where inequalities \eqref{eq:ex_1_system_3}, \eqref{eq:ex_1_system_4}, \eqref{eq:ex_1_system_6} are already implied by $\s \in \S$. 
We color the remaining three inequalities similarly to Figure~\ref{fig:ex_inertial_1},
which reports
the solution to~\eqref{eq:ex_1_system} 
(i.e., the inertial equilibrium set) in gray.
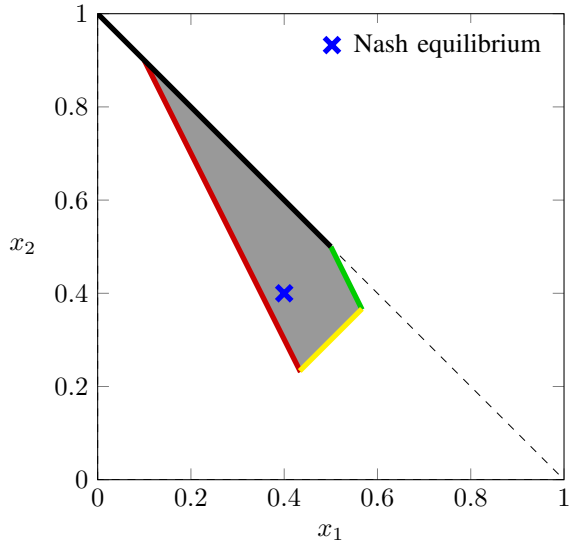
\begin{figure}[!h]
    \centering
    \newlength\figureheight
    \newlength\figurewidth
    \setlength\figureheight{6.2cm} 
	\setlength\figurewidth{6.2cm} 
%
%
\definecolor{mycolor1}{rgb}{0.85000,0.32500,0.09800}%
\begin{tikzpicture}
\begin{axis}[%
width=\figurewidth,
height=\figureheight,
scale only axis,
xmin=0,
xmax=1,
xtick={0,0.2,0.4,0.6,0.8,1},
xlabel={$\s_1$},
ymin=0,
ymax=1,
ytick={0,0.2,0.4,0.6,0.8,1},
ylabel style = {rotate=-90},
ylabel={$\s_2$},
legend style={draw = none,  align=left,align=left}
]
\addplot [dashed, line width=0.3pt, forget plot]
  table[row sep=crcr]{%
0 0\\
0 1\\
1 0\\
0 0\\
};
\addplot [color = blue, only marks, mark =x, mark options={fill=cyan, line width=2pt}, mark size=4pt]
table[row sep=crcr]{
0.4 0.4\\
};
\addlegendentry{~Nash equilibrium}

\addplot [thick,color=black!40,fill=black!40]
coordinates {
			(0,1)
            (0.1, 0.9) 
            (0.433333, 0.233333)
            (0.566667, 0.366667)
            (0.5, 0.5)
            (0.1, 0.9)};

\addplot [line width= 2pt,color=green!80!black, forget plot]
table[row sep=crcr]{ 
			0.5 0.5\\
			0.567 0.365\\          
};

\addplot [line width= 2pt,color=red!80!black, forget plot]
table[row sep=crcr]{ 
            0.1 0.9\\
            0.435 0.2312\\
};

\addplot [line width= 2pt,color=black, forget plot]
table[row sep=crcr]{ 
			0.5 0.5\\
			0 1\\          
};

\addplot [line width=2pt,color=yellow, forget plot]
table[row sep=crcr]{ 
            0.433333 0.233333\\
            0.566667 0.366667\\
};
\end{axis}
\end{tikzpicture}%
    \caption{
    The shaded region, including the thick red, yellow, green, and black lines, represents the inertial Nash equilibrium set for Example~\ref{ex:inertial_1} projected on the plane ($\s_1, \s_2$). The component $\s_3$ can be reconstructed from $\s_3 = 1-\s_1-\s_2$.
	The dashed line represents the simplex boundary, while the yellow, green and red lines describe the inequalities in \eqref{eq:ex_1_system}.
	The blue point is the unique Nash equilibrium $\bar{\s}=[0.4,0.4,0,2]$, which satisfies condition~\eqref{eq:no_improv_parallel}.}
    \label{fig:ex_inertial_1}
  \end{figure}
\end{example}
We note that the inertial equilibrium set is not a singleton. The lack of uniqueness is due to the positivity of the coefficients $\jh$. Indeed, if $\jh=0$ for all $i,j$, then the inertial equilibrium set coincides with the equilibrium set of Definition~\ref{def:Nash_parallel}, which is a singleton marked in blue in Figure~\ref{fig:ex_inertial_1}. Moreover, the inertial equilibrium set is not convex. This is due to the line joining the point $(0.1,0.9)$ to $(0,1)$ in Figure~\ref{fig:ex_inertial_1}. The points on this segment belong to the inertial equilibrium set even though they do not satisfy $\s_2\ge -2\s_1+1.1$.
This is because~\eqref{eq:ex_1_system_5} is enforced only when $\s_3>0$, whereas $\s_3=0$ on the considered segment.
The observed non-convexity of the solution set is, in a sense, structural. 
To see this, note that, by Definition 2, a point $\s \in \Spl$ is an inertial equilibrium if and only if it lies at the intersection of inequality constraints of the form $u_j(\s)-c_{ij}-u_i(\s) \le 0$; these might be non convex, even if we restrict attention to convex or concave utility functions.
\subsection{Related Work}
\label{sec:literature}
The notion of inertial equilibrium is, to the best of our knowledge, novel, due to the presence of the switching costs $c_{ij}$.
A related line of works comes from \textit{population games} \cite{sandholm2010population}. Here the focal point is the analysis and design of (continuous-time) agent dynamics that achieve an equilibrium in the sense of Definition~\ref{def:Nash_parallel}.
A particular class of dynamics is imitation dynamics. These are reminiscent of the discrete-time Algorithm~\ref{alg:better_resp} below, as agents move to more attractive actions. Different works provide local~\cite{sandholm2001potential,nachbar1990evolutionary} and global~\cite{cressman2014replicator,zino2017imitation} convergence guarantees. Rather than delving into the vast literature of population games, we observe that in all of the works there is no switching cost, i.e., $c_{ij}=0$. Thus, the literature of population games study the problem of finding an equilibrium in the sense of Definition~\ref{def:Nash_parallel} and not an equilibrium in the sense of Definition ~\ref{def:Nash_inertia}, which is the focus here.
Finally, we note that \cite{sandholm2001potential} and references therein provide convergence results to an equilibrium set, whereas we provide convergence to a point in the inertial equilibrium set. 

A more closely related equilibrium concept was proposed in the study of migration models in the seminal works  \cite{nagurney1989migration,nagurney1992human,nagurney1993human} by Nagurney. These works introduce the notion of migration equilibrium, in a way that resembles Definition~\ref{def:Nash_inertia}, but with a number of important differences.
First, the problem formulation is different.
In the migration equilibrium problem we are given a fixed initial distribution $\s^0\in\S$, with $\s^0_j$ representing the fraction of agents residing at a physical location $j$. These agents receive utility $u_j(\s^0)$. The initial distribution $\s^0$ is transformed into the final distribution $\s^1\in\S$, which is a function of the migrations $(f_{ij})_{i,j=1}^n$ (the decision variables). Each migration comes with a migration cost $c_{ij}(f_{ij})$ which is a function of the number $f_{ij}$ of agents migrating. A migration equilibrium consists of a set of migrations $(f_{ij})_{i,j=1}^n$ such that, considering the fixed initial utilities $u(\s^0)$, the migration costs $c_{ij}(f_{ij})$ and the final utilities $u(\s^1)$, no other set of migrations is more convenient. 
Second, while the better-response algorithm we will introduce in Section \ref{sec:algorithm_intertial} can be interpreted as the natural dynamics of the agents seeking an equilibrium, this is not the case for the algorithms proposed to find a migration equilibrium, which are instead VI algorithms to be carried out offline.
\section[Variational inequality reformulation]{Variational inequality reformulation}
\label{sec:VI_reformulation}

In this section we first recall that the set of equilibria defined by \eqref{eq:no_improv_parallel} can be described as the solution of a certain variational inequality. We then show that a similar result holds for the inertial equilibrium set of  \eqref{eq:no_improv}. While the former equivalence is known, the latter connection is novel and requires the careful definition of the variational inequality operator. The interest in connecting the inertial equilibrium problem with the theory of variational inequalities stems from the possibility of inheriting readily available results, such as existence of the solution, properties of the solution set, and algorithmic convergence.
Basic properties and results from the theory of variational inequalities used in this manuscript are summarized in Appendix \ref{app:VI}.

\begin{definition}[Variational inequality]
\label{def:VI}
Consider a set $\X \subseteq \R^n$ and an operator $F:\X \rightarrow \R^n$.
A point $\bar x\in\X$ is a solution of the variational inequality $\textup{VI}(\X,F)$ if
\begin{equation}
F(\bar x)^\top (x-\bar x)\ge 0, \quad \forall x\in\X.
\end{equation}
\end{definition}
\ni The variational inequality problem was first introduced in infinite dimensional spaces in~\cite{stampacchia1966}, while the finite-dimensional problem in Definition~\ref{def:VI} was identified and studied for the first time in~\cite{cottle1966thesis}. The monograph~\cite{facchinei2007finite} includes a wide range of results on VI, amongst which their connection to Nash equilibria.
\begin{proposition}[Equilibria as VI solutions, 
{\cite[Thm 2.3.2]{sandholm2010population}}]
\label{lem:parallel_VI}
A point $\bs \in \S$ is an equilibrium if and only if it is a solution of VI($\S$,$-u$),
where $u(\s)\coloneqq[u_i(\s)]\jn$.
\end{proposition}

The following theorem shows that inertial equilibria can also be described by suitable variational inequalities.
\begin{theorem}[Inertial equilibria as VI solutions]
\label{thm:equiv_VI}
A point $\bs \in \S$ is an inertial equilibrium if and only if it is a solution of VI$(\S,F)$, where
\begin{equation}
\begin{split}
\label{eq:F_def}
F(\s) &\coloneqq [F_i(\s)]_{i=1}^n, \\
F_i(\s) &\coloneqq \maxx{j \in \onen}{(u_j(\s) - u_i(\s) - c_{ij})}.
\end{split}
\end{equation}
If the utilities are continuous, the existence of an inertial equilibrium is guaranteed.
\end{theorem}

\subsection{Lack of  monotonicity}
If the operator $F$ in VI($\S,F$) is monotone (see Definition~\ref{def:monotone} in Appendix \ref{app:VI}), an inertial equilibrium can be computed efficiently using one of the many algorithms available in the literature of variational inequalities, see~\cite[Chapter 12]{facchinei2007finite}.
 On the contrary, if this is not the case, the problem is known to be intractable in general.
Since the inertial equilibrium set of Figure~\ref{fig:ex_inertial_1} is not convex, the corresponding variational inequality operator $F$ cannot be monotone (see Proposition~\ref{prop:solution_set_convex} in Appendix~\ref{app:VI}). The question is whether this observation extends to more general settings. 
In the following we provide a strong negative result showing that the variational inequality operator is \emph{non monotone} in many meaningful instances of the inertial equilibrium problem. 
\begin{theorem}[$F$ is not monotone]
\label{thm:not_MON}
Assume that for all $i\in\onen$ the function $u_i$ is Lipschitz and
that $\nabla_{\s_i}u_i(\s)<0$ for all $
\s\in\S$. If there exists a point $\hat{\s} \in\S$ which is \emph{not} an inertial equilibrium, then $F$ is not monotone in $\S$.
\end{theorem}
The theorem certifies that either every point of the simplex is an equilibrium, or $F$ is not monotone and consequently the variational inequality problem is hard.
The only technical assumption is that $\nabla_{\s_i}u_i(\s)<0$. We observe that this is the situation for many applications; indeed the condition implies that $u_i(\s)$ decreases if the number of agents on action $i$ increases, as commonly assumed in congestion problems. Moreover, the condition can be further weakened, as for the proof it suffices that $\nabla_{\s_\jj}u_\jj(\sss)<0$, only for a specific $\s^\star$ and $i^\star$ defined in Appendix~\ref{app:proofs}.

We conclude this section by pointing out that Example~\ref{ex:inertial_1} satisfies the conditions of Theorem~\ref{thm:not_MON}.
The lack of monotonicity of the corresponding operator $F$ is confirmed by the fact that  $\nabla_\s F(\s)$ is not positive semidefinite for all $\s\in\S$ (a condition equivalent to monotonicity, see Proposition \ref{prop:pos_def_generalized} in Appendix \ref{app:VI}).
Indeed, there are points where $\nabla_\s F(\s) + \nabla_\s F(\s)^\top$ is indefinite, e.g.,  $\tilde \s = [0.2, 0.2, 0.6]$, where
\begin{equation}
\label{eq:ex_parallel_1_indef}
\nabla_\s F(\tilde \s) + \nabla_\s F(\tilde \s)^\top = \begin{bmatrix} 0 & 0 & -1 \\ 0 & 0 & 0 \\ -1 & 0 & 2 \end{bmatrix}\,.
\end{equation}

\subsection{Three drawbacks of existing algorithms}
\label{subsec:parallel_Nash}
Lemma \ref{lem:parallel_is_inertial} ensures that any equilibrium is  an inertial equilibrium. Thus, one might be tempted to use an algorithm for computing an equilibrium to determine an inertial equilibrium. Unfortunately, a number of difficulties make this approach impractical. In this section we describe \emph{one} such algorithm and highlight its drawbacks in the computation of an inertial equilibrium,
which generalise to other algorithms that converge to an equilibrium. We consider the projection algorithm \cite[Alg. 12.1.1]{facchinei2007finite} for the solution of VI($\S$,$-u$), where $\s(k)$ indicates the iterate $k$ of the algorithm. Note that the projection step necessitates the presence of a central operator.

\begin{algorithm}[H]
\caption{Projection algorithm}
\label{alg:projection_parallel}
\vspace*{0.2cm}
\textbf{Initialization:} \hspace*{0.36cm} $\rho> 0$, $k=0$, $\s(0) \in \S$
\\[0.1cm]
\textbf{Iterate:}
\hspace*{1.25cm} $\s{(k+1)} = \proj{\S}\left[\s{(k)} + \rho u(\s{(k)})\right]$ \\
\hspace*{2.4cm} $k \leftarrow k+1$
\vspace*{0.2cm}
\end{algorithm}
\begin{proposition}
\label{prop:proj_converges_to_parallel}
If $u_i$ is $\Lmax$-Lipschitz for all $i$, $\rho\le 2/L$, and if there exists a concave function $\theta:\Rn\to\R$ such that $\nabla_\s \theta(\s)=u(\s)$ for all $\s\in\S$,
then Algorithm~\ref{alg:projection_parallel} converges to an equilibrium, and thus an inertial equilibrium.%
\footnote{
For this proposition to hold, we have to assume the existence of a concave function $\theta$ whose gradient matches $u(\s)$.
One such case is when the utility function $u_i$ depends only on the number of agents on action $i$, i.e. $u_i(\s)=u_i(\s_i)$ for all $i$, and is decreasing. This case covers a wide range of applications.
If no $\theta$ whose gradient matches $u(\s)$ exists, but $-u$ is monotone, one can resort to a different algorithm such as the extra-gradient algorithm \cite[Thm. 12.1.11]{facchinei2007finite}. Finally, observe that if $-u$ is strongly monotone (see \cite[Def. 2.3.1]{facchinei2007finite}), the projection algorithm converges without requiring the existence of $\theta(\s)$, see \cite[Alg. 12.1.1]{facchinei2007finite}.}
\end{proposition}
\subsection*{Three fundamental shortcomings}
In the following we analyse the behaviour of Algorithm~\ref{alg:projection_parallel} on Example~\ref{ex:inertial_1}, and use it to highlight three fundamental shortcomings of this approach.
We begin by observing that $\bs=[\bs_1,\bs_2,\bs_3]=[0.4,0.4,0.2]$ is an equilibrium, as it solves VI($\S$,$-u$), \mbox{since for all $x\in\S$}
\[
\small
\begin{bmatrix} -u_1(\bs_1) \\ -u_2(\bs_2) \\ -u_3(\bs_3) \end{bmatrix}\tr
\!\!\!\!
\left(\begin{bmatrix} \s_1 \\ \s_2 \\ \s_3 \end{bmatrix}
\! - \! 
\begin{bmatrix} \bs_1 \\ \bs_2 \\ \bs_3 \end{bmatrix} \right)  = 
\begin{bmatrix} -0.8 \\ -0.8 \\ -0.8 \end{bmatrix}\tr
\!\!\!\!
\left(\begin{bmatrix} \s_1 \\ \s_2 \\ \s_3 \end{bmatrix}
\! - \! 
\begin{bmatrix} \bs_1 \\ \bs_2 \\ \bs_3 \end{bmatrix} \right)  = 0.
\]
Additonally, $[\bs_1,\bs_2,\bs_3]=[0.4,0.4,0.2]$
is the \emph{unique} solution of VI($\S$,$-u$) and thus the unique equilibrium (see~\cite[Thm. 2.3.3]{facchinei2007finite}).
This is consistent with  
Lemma~\ref{lem:parallel_is_inertial} and Figure~\ref{fig:ex_inertial_1} where the equilibrium point $\bs$ belongs to the inertial equilibrium set.
Thanks to Proposition~\ref{prop:proj_converges_to_parallel}, Algorithm~\ref{alg:projection_parallel} converges to $\bs$
 ($L=1$ for the utilities in~\eqref{eq:ex_1_system}, so that we have to select $\rho<2$).
With the choice of $\rho=1$, it is immediate to verify that Algorithm~\ref{alg:projection_parallel} converges in one iteration for any initial condition $\s(0)$.
  
In the following we consider two cases: i) the case in which $\s(0)$ is neither an inertial equilibrium nor an equilibrium; ii) the case in which $\s(0)$ is an inertial equilibrium, but not an equilibrium.
Case i): consider $\s(0) = [0.4, 0.2, 0.4]$. The point $\s(0)$ is not an inertial equilibrium (and thus not an equilibrium), because
$\s_3(0)>0$ and $u_3(\s(0))=1-0.4=0.6<0.7=0.8-0.1=u_1(\s(0))-c_{31}$.
The first iteration of Algorithm~\ref{alg:projection_parallel} amounts to a mass of $0.2$ being moved from action $i=3$ to action $i=2$. 
Nevertheless, we observe that agents selecting action $i=3$ are not interested in moving to action $i=2$. Indeed
$u_3(\s(0))=0.6 \ge -0.2=u_2(\s(0))-c_{32}$, so the switch from $i=3$ to $i=2$ is detrimental for the agents performing it. 
Case ii): Consider $\s(0)=[0.4, 0.3, 0.3]$, and note that $\s(0)$ is already an inertial equilibrium.
Nonetheless, the first iteration of Algorithm~\ref{alg:projection_parallel} forces a mass of $0.1$ to move from action $3$ to $2$.

The drawbacks of Algorithm~\ref{alg:projection_parallel} are summarized next:%
\begin{enumerate}[i)]
\item Agents are forced to switch action even when such switch is detrimental to their well being.
\item Agents are forced to switch action even if already at an inertial equilibrium.
\item The projection step necessitates the presence of a central operator. Such operator requires information not only on the utilities $u_i(\s(k))$ for all $i$, but also on $\s(k)$.
\end{enumerate}
In the next section we overcome these issues and present a natural dynamics that i) provably converges to an inertial equilibrium, ii) respects the agent's strategic nature, and iii) requires limited coordination.
\section{A better-response algorithm}
\label{sec:algorithm_intertial}
We begin by introducing the definition of the envy set.
\begin{definition}[Envy set]
\label{def:envyset}
Given $\s \in \S$, for each $i$ such that $\s_i > 0$, we define the envy set of $i$ as
\begin{equation}
\Nienvy(\s) \defeq \left\{ j \in \Niout \text{ s.t. } u_i(\s) < u_j(\s) - c_{ij} \right\},
\end{equation}
whereas for $i$ such that $\s_i = 0$, we define $\Nienvy(\s) = \emptyset$.
\end{definition}
\noindent 
Informally, the envy set $\Nienvy(\s)$ contains all the actions $j$ to which agents currently selecting action $i$ would rather move to.
The following fact immediately follows from Definition \ref{def:envyset} and Definition~\ref{def:Nash_inertia} of inertial equilibrium.	

\begin{proposition}
\label{prop:equiv_out_envy}
A point $\bs\in\S$ is an inertial equilibrium if and only if
$
\bs \in \S \text{ and } \Nienvy(\bs) = \emptyset, \; \text{for all} \; i \in \onen.	
$
\end{proposition}
\vspace*{2mm}
The proposed Algorithm \ref{alg:better_resp} involves  a single, intuitive step. At iteration $k$, let $x(k)\in\S$ denote the distribution of the agents on the resources. For every action $i$, a mass $\xijk\in[0, \s_i(k)]$ is moved from action $i$ to some other action $j\in\Nienvy(\s(k))$, that is, the movement takes place only if the alternative action $j$ is attractive for agents currently selecting action $i$. This simple dynamics is described in Algorithm~\ref{alg:better_resp}, where we denote with $u_i(k)=u_i(\s(k))$, $\Nenvy_i(k) = \Nenvy_i(\s(k))$ for brevity.
\begin{algorithm}[H]
\caption{Better-response algorithm}
\label{alg:better_resp}

\textbf{Initialization}: \hspace*{0.5cm} $k =0$, $\s(0) \in \S$ 

\vspace{0.1cm}
\textbf{Iterate}: \hspace*{1.42cm} $\Delta x(k)\gets 0$\\
\hspace*{2.65cm} \textbf{repeat} for all $i$, $j\in\Nenvy_i(k)$\\
\hspace*{3.3cm} choose $\xijk\in[0, \s_i(k)]$ \\
\hspace*{3.3cm} $\Delta\s_i(k) \leftarrow \Delta\s_i(k) - \xijk$, \\
\hspace*{3.3cm} $\Delta\s_j(k) \leftarrow \Delta\s_j(k) + \xijk$, \\
\hspace*{2.65cm} \textbf{end repeat} \\
\hspace*{2.65cm} $x(k+1)\gets x(k)+\Delta x(k)$\\
\hspace*{2.65cm} $k \leftarrow k+1$
\end{algorithm}

The agents' dynamics presented in Algorithm \ref{alg:better_resp} is fully specified once we define the mass $\xijk$ moving from action $i$ to $j\in\Nenvy_j(k)$ as a function of $\s_i(k)$ and $\s_j(k)$. At this stage we rather not give a particular expression to $\xijk$, as the convergence of Algorithm~\ref{alg:better_resp} is guaranteed under very weak conditions and different choices of $\xijk$. 
A possible modelling assumption sees agents moving from a less attractive action $i$ to a more favourable action $j\in \Nienvy(k)$ \emph{independently} from the value of the utility $u_j(k)$. For instance, this can be achieved by setting $\xijk = \beta \s_i(k)$ with $\beta>0$.
A different modelling assumption entails agents being responsive to the level of the utility $u_j(k)$ over all $j\in \Nienvy(k)$, and thus redistributing themselves based on the perceived gain. Both these cases (and many more) are covered by Theorem \ref{thm:convergence_alg_distr}.

We observe that Algorithm~\ref{alg:better_resp} does not present any of the issues encountered with the use of Algorithm~\ref{alg:projection_parallel}.
First, agents switch action only if the switch is convenient. Second, no agent moves if the current allocation is an inertial equilibrium. Third, there is no need for a central operator, and each agent requires information only regarding the other actions' utilities $u(\s(k))$.
As a consequence, Algorithm~\ref{alg:better_resp} can be interpreted as the \emph{natural dynamics} of agents switching to a more favourable action whenever one is available.
Finally, agents are not limited to moving to the best alternative action (as in \textit{best-response dynamics}), but can instead choose any action providing a better net utility (hence the term \textit{better-response dynamics}).

\begin{theorem}[Convergence of Algorithm \ref{alg:better_resp}]
\label{thm:convergence_alg_distr}
Assume that
\begin{itemize}
	\item[-] for each $i \in \onen$ the utility $u_i$ depends only on $\s_i$, that $u_i$ is non-increasing and $L$-Lipschitz.
	\item[-] there exists $\cmin>0$ such that $\jh\ge\cmin$ for all $i\neq j$ with $i,j\in\onen$.
	\item[-] there exist $0< \tau\le 1$, and $\varepsilon>0$ such that at each iteration $k \in \mb{N}$, $\xijk\ge0$ for all $i\in\onen$, $j\in \Nienvy(\s_k)$, and
		\begin{subequations}
		\begin{align}
		&\hspace*{-6mm}\tau \s_i(k) 
		\!\le\!\!\!\!
		\sum_{j\in \Nenvy_i(k)} 
		\!\!\!\!
		\xijk \le  \s_i(k), & & \hspace*{-4mm} \, i\in\{1,\dots,n\}, \;\label{eq:bound_lower} \\
		&\hspace*{4.9mm}\sum_{i : j\in\Nenvy_i(k)} \hspace*{-0.4cm} \xijk \le \frac{\cmin}{\Lmax} - \varepsilon, & & \hspace*{-3mm} \, j\in\{1,\dots,n\}.\ \label{eq:bound_upper}
		\end{align}
		\label{eq:bounds}
		\end{subequations}
\end{itemize}
\noindent 
Then $\s(k)$ in Algorithm~\ref{alg:better_resp} converges to an inertial equilibrium $\bs$.
If additionally $\bs \in\Spl$, then the algorithm terminates in a finite number of steps.
\end{theorem}

The first assumption is typical of many congestion-like problems.
The second assumption is technical, and requires the switching costs between different actions to be strictly positive.  
With respect to the third assumption, the requirement on the right hand side of \eqref{eq:bound_lower} together with the condition $\xijk\ge0$ for all $i\in\onen$, $j\in \Nienvy(\s_k)$, is needed to ensure that $x(k)$ remains in the simplex. Thus, the only non-trivial constraint imposed on $\xijk$ is that on the left hand side of \eqref{eq:bound_lower}, and that of \eqref{eq:bound_upper}; these are discussed in detail in Remark 1 below. Finally, we note that the proof of Theorem~\ref{thm:convergence_alg_distr} does not require the agents to move synchronously. As a consequence, an asynchronous implementation of Algorithm \ref{alg:better_resp} is also guaranteed to converge.

\begin{remark}[Tightness of conditions \eqref{eq:bound_lower} and \eqref{eq:bound_upper}]
Condition~\eqref{eq:bound_lower} is a mild requirement.
It merely asks for a minimum proportion of agents to move from their current unfavourable action to a better one.
Equation~\eqref{eq:bound_upper}, on the other hand, requires the switching to happen sufficiently slowly. Without this condition, the algorithm may not converge, as shown with the following example.
Consider $n=2$, $u_1(\s_1)=1-\s_1$, $u_2(\s_2)=1-\s_2$, $c_{12}=c_{21}=0.5$, and note that $\cmin / \Lmax = 0.5$.
Take $\delta>0$ small enough and initial condition $\s_1(0)=0.75+\delta/2$, $\s_2(0)=0.25-\delta/2$.
Since $u_1(0)=0.25-\delta/2$ and $u_2(0)=0.75+\delta/2$, then $\s(0)$ is not an inertial equilibrium. Assume that, as a consequence, $0.5+\delta>\cmin / \Lmax $ units of mass move from action $1$ to action $2$, resulting in $x_1(1)=0.25-\delta/2$, $x_2(1)=0.75+\delta/2$, and thus $u_1(1)=0.75+\delta/2$, $u_2(1)=0.25-\delta/2$, so $x(1)$ is not an inertial equilibrium either.
A repeated transfer of $0.5+\delta$ mass from the action which is worse-off to the one which is better-off results in
$x(2k)=x(0)$ and $x(2k+1)=x(1)$. Thus, a slight violation of \eqref{eq:bound_upper} brakes the convergence of Algorithm~\ref{alg:better_resp}.
\end{remark}
\section{Extensions}
\label{sec:extensions}

We present three modifications of the inertial equilibrium problem, and highlight how the results can be adapted.\\[-3mm]

\ni{\bf Non-engaging agents.}
With the current Definition \ref{def:Nash_inertia} all the agents are forced to engage, i.e. to choose one of the actions in $\onen$.
Let us now consider an extra action labeled $e$, so that the extended actions set it $\{1,\dots,n,e\}$.
Set $c_{je}=c_{ej}=0$ for all $j\in\onen$ and $u_e(\s)$ 
as some constant value representing, for instance, the utility perceived when not participating in the game. Within this setup, an agent that does not engage in the game at time $k=0$, will revise his decision at every time-step $k\ge1$, and will rejoin whenever more favourable actions appear.
For example, in the ride-hailing application presented in Section \ref{sec:application}, action $e$ could represent electing to temporarily not work as a driver.
\\[-3mm]

\ni {\bf Atomic agents with discrete action set.}
Instead of a continuum of agents,
one could consider a finite number $\N$ of atomic agents.
Each agent possesses unitary mass and can choose only one of the actions $\{1,\dots,n\}$.
The utility $u_j$ is then a function of how agents distribute themselves over the actions.
The definition of inertial equilibrium requires that no agent $i\in\oneN$ has an incentive to switch action,
considering the utilities of the alternative actions and the corresponding switching costs.
The model with a continuum of agents studied above represents, in a sense, the limiting case obtained as the number of agents $\N$ grows.
Since the action space is discrete,
the reformulation as a VI is not possible.
Nonetheless, one can formulate Algorithm~\ref{alg:better_resp}
by letting an agent $i$ switch to an arbitrary action whenever such action is attractive.
Convergence is guaranteed upon
substituting the expression $\sum_{j\in \Nenvy_i(k)} 
		\xijk$ in \eqref{eq:bound_lower} and \eqref{eq:bound_upper} with the number of agents that move at the same time.\\[-3mm]

\ni {\bf Multi-class inertial equilibrium.} The concept of inertial equilibrium relies on the idea that each agent perceives the same utility $u_j$ and
the same switching costs $c_{ij}$.
This assumption can be relaxed by introducing different agents' classes. Let $A$ be the total number of classes, and $\sja$ be the mass of agents belonging to class $\alpha\in A$ which choose action $j$.
We denote $\sj = \suma \sja$ and $\sa = \{\sja\}\jn$.
\begin{definition}
\label{def:Nash_inertia_multiclass}
Consider utilities $u_i^\a:\R^{n}_{\ge 0}\to\R$, switching costs $c_{ij}^\a \ge 0$ and masses $\gamma^\a > 0$, with $i,j \in\onen$, $\a\in\oneA$.
The vector $\bs = [\bs^1,\dots, \bs^A] \in \RnA$ is a \textit{multi-class inertial equilibrium} if $\bs \ge \zeros[n A]$,
$\ones[n]\tr \bs^\a = \gamma^\a$ for all $\a$, and
\begin{equation}
\bs^\a_{i}> 0 \implies u^\a_i(\bs_{r}) \ge u^\a_j(\bs_{r}) - c^\a_{ij}, \quad \forall \, j\in\onen,
\label{eq:def_parallel}
\end{equation}
for all $i \in \onen$ and $\a \in \oneA$, where the vector $\bs_{r}\coloneqq\suma \bs^\a$.
\end{definition}
\ni
Note that even though different classes might perceive different utilities at the same action $i$, each of these utilities is a function of the sole distribution of the agents on the actions i.e. of the reduced variable $\s_{r}$.
This is indeed what couples the different classes together.
Upon redefining
$\S = \tilde\S^1\times \dots \times \tilde\S^A \subset \RnA$ as the Cartesian product of the weighted simplexes $\tilde\S^\alpha=\{\s^\alpha\in\mb{R}^n_{\ge0},~ \ones[n]\tr \s^\a = \gamma^\a\}$,
one can redefine
$F: \S \rightarrow \Rpl^{nA}$,
where
\begin{equation}
\begin{aligned}
\label{eq:F_def_multi}
&F(\s) = [[F_j^\a(\s)]_{\a=1}^A]_{j=1}^n, \\
&F_j^\a(\s) = \maxx{h \in \onen}{\left(u_h^\a\left(\suma \s^\a\right) - u_j^\a\left(\suma \s^\a\right) - c^\a_{jh}\right)}.
\end{aligned}
\end{equation}
Using a straightforward extension of the proof of Theorem~\ref{thm:equiv_VI},
one can show that the set of multi-class inertial equilibria coincides with the solution set of
VI$(\mc{S},F)$.
Theorem~\ref{thm:not_MON} about lack of monotonicity
also extends to the multi-class case. 
Finally, Algorithm~\ref{alg:better_resp} can also be modified appropriately to account for the presence of multiple classes, and a similar convergence result to that of Theorem~\ref{thm:convergence_alg_distr} follows. 
\section{Application: area coverage for taxi drivers}
\label{sec:application}
In this section we apply the theory developed to the problem of area coverage for taxi drivers. Understanding the spatial behavior of taxi drivers has attracted the interest of the transportation community~\cite{Castro2012,li2011hunting}, as it allows to infer information for diverse scopes, including land-use classification~\cite{pan2013land} and analysis of collective behaviour of a city's population~\cite{castro2013taxi}.

We focus on the urban area of Hong Kong, as the work~\cite{wong2014modelling} provides relevant data for our model. The authors of \cite{wong2014modelling} divide the region of interest into $n=18$ neighborhoods, which represent the resources in our game.
We assume that a taxi driver in neighborhood $i$ enjoys the utility $u_i(x_i)$, depending on the fraction $x_i$ of taxi drivers covering the same neighborhood.
We aim at determining an equilibrium distribution of the drivers, across the different neighborhoods of the urban area.
The problem can be described through the introduction of an undirected graph, where the nodes represent the neighborhoods. We construct an edge from $i$ to $j$ (and from $j$ to $i$) if and only if the two neighborhoods are adjacent. The cost $c_{ij}$ is taken as the fuel cost of a trip from $i$ to $j$ according to~\cite{ViaMichelin} and $c_{ij} = c_{ji}$,
with the fuel cost set as extremely high for non-adjacent neighborhoods, so that movement cannot occur between those.
A taxi driver stationing in node $i$ experiences a utility $u_i(x_i)$ describing his revenue minus the costs. This takes the form of 
\begin{equation}
u_i(x_i) = \alpha_i v_i(x_i) - (1-v_i(x_i)) \beta
\end{equation}
where $\alpha_i$
is the average profit per trip starting from location $i$
(ranging from 30 to 140 HK\$ according to \cite{wong2014modelling}),
$\beta = 6.34$ HK\$ is the operational cost of vacant taxi trips inclusive of fuel costs, rental costs and toll charges associated with the trips.
The function $v_i(x_i)$ describes the percentage of the time a taxi is occupied and according to~\cite[Eq. (1)]{buchholz2015spatial} is modeled~by 
\begin{equation}
v_i(x_i) = 1-\left(\frac{x_i}{1+x_i}\right)^{p_i},
\end{equation}
where $p_i > 1$ is the number of passengers requesting a taxi at node $i$. We select $p_i$ to be proportional to the values in Figure 3 of~\cite{wong2014modelling}.
Note that $v_i(0) = 1$ and $\lim_{x_i \to \infty} v_i(x_i) = 0$, as one would expect.
Moreover, through simple algebraic manipulations, $u_i(x_i)$ can be shown to be decreasing for $x_i \ge 0$ and to have Lipschitz constant
$
L_i = 4(\alpha_i-\beta)p_i\frac{(p_i-1)^{p_i-1}}{(p_i+1)^{p_i+1}}.
$
In our numerical study we compare the projection algorithm (Algorithm~\ref{alg:projection_parallel}) with the better-response algorithm proposed in Algorithm~\ref{alg:better_resp}, with stopping criterion
$\| x(k+1)-x(k)\| \le 10^{-6}$, and equal neighbour redistribution function $\xijk=\tau x_i(k)$, $j\in\Nienvy(k)$.
For Algorithm~\ref{alg:projection_parallel}, $\rho$ is chosen slightly smaller than $1/ \Lmax = 1.7 \cdot 10^{-3}$ as required to achieve convergence by Proposition~\ref{prop:proj_converges_to_parallel}. Similarly, $\tau$ is chosen slightly smaller than $c_\tmin/ (\Lmax \gamma) = 1.4 \cdot 10^{-4}$
 in accordance to the requirement of Theorem~\ref{thm:convergence_alg_distr}.
Table~\ref{table:iterations} (top) shows a comparison in terms of iterations needed to reach convergence by both algorithms.
Note that a single iteration of Algorithm~\ref{alg:projection_parallel} is more costly than one of Algorithm~\ref{alg:better_resp}. Indeed, Algorithm~\ref{alg:projection_parallel} requires the computation of a projection step, while Algorithm~\ref{alg:better_resp} requires simple addition and multiplication operations.

\begin{figure}[t!]
	\begin{center}
		\includegraphics[width = 0.95\columnwidth]{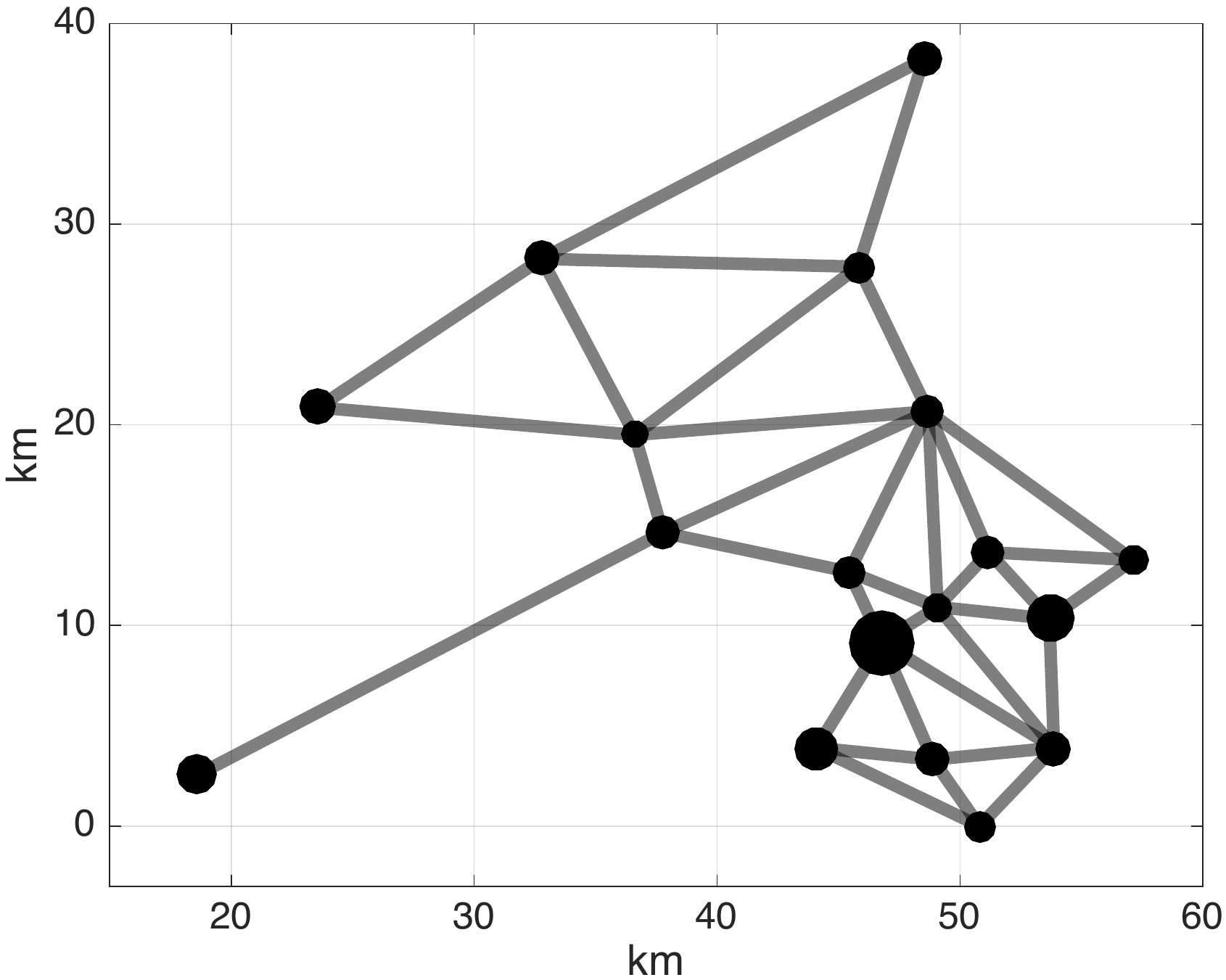}
	\end{center}
	\caption{The equilibrium $\bar x$ achieved by Algorithm~\ref{alg:better_resp} with initial condition $x(0) = \ones[n]/\gamma$.
	The radius of each node is proportional to its utility $u_i(\bar x_i)$, while the thickness of edge $(i,j)$ is proportional to the corresponding cost $c_{ij}$.}
	\label{fig:equilibrium_hong_kong}
\end{figure}

We note that the number of required iterations to reach convergence is rather high, due to the small values of $\rho$ and $\tau$ imposed by the theoretical bounds of Proposition~\ref{prop:proj_converges_to_parallel} and Theorem~\ref{thm:convergence_alg_distr}.
For this reason we perform another simulation with the values $\rho = \tau = 10^{-2}$, which provide no theoretical guarantees of convergence.
Nonetheless, both algorithms converge in $100$ different repetitions with random initial conditions. The number of iterations is reported in Table~\ref{table:iterations} (bottom) and is considerably smaller than those in Table~\ref{table:iterations} (top). Moreover, Algorithm~\ref{alg:projection_parallel} outperforms Algorithm~\ref{alg:better_resp} in the first case, but the viceversa happens in the second case.
Finally, Figure \ref{fig:equilibrium_hong_kong} shows the steady state distribution of taxi drivers across the $n$ neighbourhood of Hong Kong, with initial condition $x(0)=\ones[n]/\gamma$.

\begin{table}[H]
\begin{center}
\begin{tabular}{ccc}
{\bf Algorithm} & {\bf \# iterations mean } & {\bf \# iterations St. Dev.} \\ 
\specialrule{.1em}{.2em}{.2em} 
Alg.~\ref{alg:projection_parallel}, $\rho= 10^{-3}$& $\num{26311}$ & $\num{3329}$ \\
\specialrule{.05em}{.2em}{.2em} 
Alg.~\ref{alg:better_resp}, $\tau= 10^{-4}$ & $\num{152856}$ & $\num{9130}$ \\
\specialrule{.1em}{.2em}{.2em} 
Alg.~\ref{alg:projection_parallel}, $\rho=10^{-2}$& $\num{12672}$ & $\num{3472}$ \\
\specialrule{.05em}{.2em}{.2em} 
Alg.~\ref{alg:better_resp}, $\tau= 10^{-2}$ & $\num{2168}$ & $\num{125}$ \\ 
\specialrule{.1em}{.2em}{.2em} 
\end{tabular}
\end{center}
\vspace*{3mm}
\caption{Number of iterations needed to reach convergence with 
$\rho=1.5 \cdot 10^{-3}$ , $\tau= 10^{-4}$ (top), and $\tau=\rho=10^{-2}$ (bottom). We report mean and standard deviation for $100$ repetitions of the two algorithms, starting from random initial conditions in the simplex.}
\label{table:iterations}
\end{table}
\section{Conclusions}
\label{subsec:conclusions}
We proposed the novel notion of inertial Nash equilibrium to model  the cost incurred by agents when switching to an alternative action. 
While the set of inertial Nash equilibria can be characterized by means of a suitable variational inequality, the resulting operator is often non monotone. Thus, we proposed a natural dynamics that is distributed, and provably converges to an inertial Nash equilibrium.
As future research direction, it would be interesting to provide convergence rate guarantees for Algorithm~\ref{alg:better_resp}, and more broadly to extend the notion of inertial equilibrium beyond the framework of population games.

\appendices
\section{Preliminaries on variational inequalities}
\label{app:VI}
\label{sec:VI}
In the following we present those result on the theory of variational inequality that are used to characterize the equilibrium concepts introduced in Section~\ref{sec:inertial_Nash}.
\begin{proposition}[{\cite[Prop. 2.3.3]{facchinei2007finite}}]
\label{prop:existence_VI_sol}
Let $\X \subset \Rn$ be a compact, convex set and $F: \X\to\Rn$ be continuous.
Then \textup{VI}($\X$,$F$) admits at least one solution.
\end{proposition}

The next proposition introduces the KKT system of a variational inequality,
which is analogous to the KKT system of an optimization program.

\begin{proposition}[{\cite[Prop. 1.3.4]{facchinei2007finite}}]
\label{prop:KKT_VI}
Assume that the set $\X$
can be described as $\X = \{ x \in \R^n \, \vert \, g(x) \le \zeros[m], h(x) = \zeros[p] \},$ and
that it
satisfies Slater's constraint qualification in~\cite[eq. (5.27)]{boyd:vandenberghe}.
Then $\bx$ solves VI($\X$,$F$) if and only if
there exist $\blambda$ and $\bmu$ such that $(\bx,\blambda,\bmu)$ solves the KKT system~\eqref{eq:KKT_VI}
\begin{subequations}
\label{eq:KKT_VI}
\begin{align}
&F(x) + \nabla_x g(x) \lambda + \nabla_x h(x) \mu = \zeros[n] \label{eq:KKT_VI_a} \\
&\zeros[m] \le \lambda \perp g(x) \le \zeros[m] 				\label{eq:KKT_VI_b} \\
&h(x) = \zeros[p].												\label{eq:KKT_VI_c}
\end{align}
\end{subequations}
\end{proposition}
We next recall the notion of monotonicity,
which is a sufficient condition for convergence of a plethora of VI algorithms,
see~\cite[Chapter 12]{facchinei2007finite}.
\begin{definition}[Monotonicity]
\label{def:monotone}
An operator $F: \X\subseteq \Rn \to \Rn$ is monotone
if for all $x,y \in \X$.
\[
(F(x)-F(y))^\top(x-y) \ge 0,
\]
\end{definition}

\begin{proposition}
{\cite[Prop. 2.1]{schaible1996generalized}}
\label{prop:pos_def_generalized}
Let $\X\subseteq\Rn$ be convex.
An operator $F$ is monotone in $\X$ if and only if
for every $x \in \X$ each generalized Jacobian $\phi \in \partial F(\s)$ is positive semi-definite.
\end{proposition}
The definition of generalized Jacobian $\partial F(\s)$ can be found in~\cite[Definition 2.6.1]{clarke1990optimization};
we do not report it here because
for our scope it suffices to know that if $F$ is differentiable in $\s$,
then the generalized Jacobian coincides with the Jacobian, i.e., $\partial F(\s) = \{\nabla_\s F(\s)\}$, with positive-definite interpreted as 
$(\nabla_x F(x) + \nabla_x F(x)\tr)/2 \succeq 0$.
We conclude this section with a result on the convexity of the VI solution set.
\begin{proposition}[{\cite[Thm. 2.3.5]{facchinei2007finite}}]
\label{prop:solution_set_convex}
Let $\X\subseteq \Rn$ be closed, convex and $F:\X\to\Rn$ be continuous and monotone.
Then the solution set of VI$(\X,F)$ is convex.
\end{proposition}
\section{Proofs}
\label{app:proofs}
\subsection*{\bf Proof of Theorem \ref{thm:equiv_VI}}
\begin{proof}
The proof consists in showing that the KKT system of VI($\S,F$) is equivalent to
Definition~\ref{def:Nash_inertia} of inertial Nash.
Since the set $\S$ satisfies Slater's constraint qualification,
by Proposition~\ref{prop:KKT_VI}
VI($\S,F$) is equivalent to its KKT system
\begin{subequations}
\label{eq:basic_model_vi_kkt}
\begin{align}
&F(\s) + \mu\ones[n] - \lambda = \zeros[n]	\label{eq:basic_model_vi_kkt_a}\\
&\zeros[m] \le \lambda \perp x \ge \zeros[m] 	\label{eq:basic_model_vi_kkt_b}\\
& \ones[n]\tr \s = 1\label{eq:basic_model_vi_kkt_c}
\end{align}
\end{subequations}
where $\mu \in \R$ is the dual variable corresponding to the constraint $\ones[n]^\top \s = 1$ and
$\lambda \in \R^n$ is the dual variable corresponding to the constraint $\s \ge \zeros[n]$.
The system~\eqref{eq:basic_model_vi_kkt}
can be compactly rewritten as
\begin{subequations}
\label{eq:basic_model_KKT_compact_ab}
\begin{align}
&\zeros[n] \leq \mu\ones[n] + F(\s) \perp \s \geq \zeros[n], \label{eq:basic_model_KKT_compact_a} \\
& \ones[n]^\top \s = 1. \label{eq:basic_model_KKT_compact_b}
\end{align}
\end{subequations}
Observe that for any $\s\in\S$ there exists $\jj \in \onen$ such that $F_\jj(\s)=0$. Indeed, setting 
\[
\jj \in \argmaxx{i\in \onen}{u_i(\s)},
\]
gives $F_{\jj}(\s)=0$ by the definition of $F$ in~\eqref{eq:F_def}.

It follows that $\mu < 0$ is not possible, otherwise the non-negativity condition on $\mu\ones[n] + F(\s)$ is violated.
Moreover, since $F(\s) \geq \zeros[n]$, $\mu > 0$ is not possible,
as by~\eqref{eq:basic_model_KKT_compact_a} this would imply $\s = \zeros[n]$ thus violating~\eqref{eq:basic_model_KKT_compact_b}.
We can conclude that $\mu=0$ and~\eqref{eq:basic_model_KKT_compact_ab} becomes
\begin{subequations}
\label{eq:basic_model_KKT_compact_cd}
	\begin{align}
	&\zeros[n] \leq F(\s) \perp \s \geq \zeros[n], \label{eq:basic_model_KKT_compact_c} \\
	& \ones[n]^\top \s = 1. \label{eq:basic_model_KKT_compact_d}
	\end{align}
	\end{subequations}
The system~\eqref{eq:basic_model_KKT_compact_cd} is equivalent to
\begin{align}
&\s \in \S, \; \text{and} \\
&\sj > 0 \underset{\text{\eqref{eq:basic_model_KKT_compact_c}}}{\Rightarrow}
u_i(\s) \ge u_j(\s)-\jh, \; \forall \, i,j \in \onen.
\end{align}
which coincides with Definition~\ref{def:Nash_inertia}.

Existence of an inertial equilibrium follows readily from Proposition~\ref{prop:existence_VI_sol} on the existence of VI solutions. The continuity of the VI operator therein required is satisfied because $F$ is the point-wise maximum of continuous functions.
\end{proof}

\subsection*{\bf Proof of Theorem \ref{thm:not_MON}}
\begin{proof}
The proof is composed of four parts.

\ni 1) We first show that there exists $\ss\in \Spl$ such that $\ss$ is not an inertial equilibrium
(by assumption $\hat x$ belongs to $\S$ and not necessarily to $\Spl$).\\ For the sake of contradiction,
assume that each $\s\in\Spl$ is an inertial equilibrium.
Since $\hat\s$ belongs to the closure of $\Spl$,
we can construct a sequence $(\s(m))_{m=1}^\infty\in\Spl$ such that $\lim_{m\to\infty}\s(m)=\hat\s$.
Since each $\s(m)$ is an inertial equilibrium and it is positive,
then for all $i,j$ it holds $u_i(\s(m))\ge u_j(\s(m))-\jh$.
Taking the limit and exploiting continuity of $\{u_i\}\jn$ we obtain
\begin{equation}
\begin{aligned}
&\lim_{m\to\infty} u_i(\s(m)) \ge \lim_{m\to\infty} u_j(\s(m))-\jh, \\
&\Leftrightarrow
u_i(\hat \s) \ge u_j(\hat \s)-\jh,
\end{aligned}
\label{eq:continuity_argument}
\end{equation}
for all $j,h \in \onen$,
hence $\hat{\s}$ is an inertial equilibrium, against the assumption.

\ni
2)
After establishing the existence of $\ss\in\Spl$ which is not an inertial equilibrium, we now show that there exists an open ball $\B_\epss(\ss)$ centered around $\ss$ of radius $\epss>0$ such that none of the points in $\B_\epss(\ss)\cap \Spl$ is an inertial equilibrium. Let us reason again for the sake of contradiction. If for each $\varepsilon>0$ there exists an inertial equilibrium in $\B_\varepsilon(\ss)\cap \Spl$, then we can construct a sequence of inertial  equilibria converging to $\ss$. With the same continuity argument used in~\eqref{eq:continuity_argument}, we can conclude that $\ss$ is an inertial equilibrium, which is false by assumption. This demonstrates the existence of $\epss>0$ such that none of the points in $\B_\epss(\ss)\cap\Spl$ is an inertial equilibrium. By Rademacher's theorem~\cite[Thm. 2.14]{ambrosio2000functions}, Lipschitzianity of $\{u_i\}\jn$ guarantees\footnote{\label{foot:Rademacher}Rademacher's theorem assumes $F$ to be defined on an open subset of $\Rn$, but $\Spl$ is not open in $\Rn$. Indeed, one just needs to define $F$ on the $n-1$ dimensional open set $\{\s\in\R^{n-1}_{>0} \vert \ones[n-1]\tr\s<1 \}$, by using $\s_n=1-\sum_{j=1}^{n-1}\s_j$ and then apply the Rademacher's Theorem to conclude existence of a differentiable point in $\{\s\in\R^{n-1}_{>0} \vert \ones[n-1]\tr\s<1 \}$ which implies existence of a differentiable point in the original $\Spl$.} existence of $\sss\in\B_\epss(\ss)\cap\Spl$ such that $F$ is differentiable at $\sss$.

\ni
3)
The previous part guarantees differentiability of $F$ at a point $\sss\in\Spl$ which is not an inertial equilibrium.
This third part is dedicated to showing that there exist $\jj,\hh\in\onen$ such that
$\jj\in\A(\hh,\sss)$ and $\A(\jj,\sss)=\{\jj\}$,
where we denote
\begin{equation}
\A(k,\s)\defeq\argmaxx{\ell \in\onen}{\left\{u_\ell(\s)-u_k(\s)-c_{k\ell}\right\}}.
\label{eq:A_argmax}
\end{equation}
Since $\sss$ is not an inertial equilibrium,
then there exist $\ell_1,\ell_2$ such that
\begin{equation}
u_{\ell_1}(\sss) < u_{\ell_2}(\sss) - c_{\ell_1 \ell_2}.
\label{eq:proof_not_MON_temp_b}
\end{equation}
Condition~\eqref{eq:proof_not_MON_temp_b} is equivalent to
$\ell_2\in\A(\ell_1,\sss)$ and $\ell_1\notin \A(\ell_1,\sss)$.
If $\A(\ell_2,\sss) = \{\ell_2\}$ then the statement is proven with $\hh=\ell_1,\jj=\ell_2$,
otherwise there exists $\ell_3 \in \A(\ell_2,\sss) \backslash \{\ell_2\}$.
Note that it cannot be $\ell_3 = \ell_1$, because this means
\begin{equation}
u_{\ell_2}(\sss) \le u_{\ell_1}(\sss) - c_{\ell_2 \ell_1},
\label{eq:proof_not_MON_temp_a}
\end{equation}
which together with~\eqref{eq:proof_not_MON_temp_b} results in
\begin{equation}
u_{\ell_1}(\sss) < u_{\ell_1}(\sss) - c_{\ell_2 \ell_1} - c_{\ell_1 \ell_2},
\end{equation}
which is not possible, because $c_{\ell_1 \ell_2}, c_{\ell_2 \ell_1} \ge 0$ by assumption.
Hence we established that $\ell_3 \neq \ell_1$.
If $\A(\ell_3,\sss) = \{\ell_3\}$ then the statement is proven with $\hh=\ell_2,\jj=\ell_3$,
otherwise there exists $\ell_4 \notin \{\ell_1,\ell_2,\ell_3\}$ such that
$\ell_4\in\A(\ell_3,\sss)$.
Since there are only $n$ different actions,
by continuing the chain of reasoning we conclude that there exists $k \in \{2,\dots,n\}$ such that
$\ell_{k} \in \A(\ell_{k-1},\sss)$ and $\A(\ell_k,\sss)=\{\ell_{k}\}$, thus proving the statement with $\hh=\ell_{k-1}$ and $\jj=\ell_k$.\\
We now proceed to show that not only $\jj\in\A(\hh,\sss)$, but actually $\A(\hh,\sss)=\{\jj\}$.
For the sake of contradiction, assume that there exists $\ell\neq\jj$ such that $\ell\in\A(\hh,\sss)$.
This means that $F_\hh(\sss)=u_\jj(\sss)-u_\hh(\sss)-c_{\hh\jj} = u_\ell(\sss)-u_\hh(\sss)-c_{\hh\ell}$.
Then consider the vector of the canonical basis $\mathbf{e}_\jj\in\Rn$ and compute
\begin{equation}
\begin{aligned}
&\lim_{t\to 0^+} \frac{F_\hh(\sss+t\mathbf{e}_\jj)-F_\hh(\sss)}{t} = \\
&\lim_{t\to 0^+} \frac{[u_\ell(\sss) \! - \! u_\hh(\sss) \! - \! c_{\hh \ell}] \! - \! [u_\ell(\sss) \! - \! u_\hh(\sss) \! - \! c_{\hh \ell}]}{t} \\
&= 0,
\end{aligned}
\label{eq:limit_incremental_ratio_a}
\end{equation}
where the first equality holds because for $t>0$ we have
\begin{align}
&u_\jj(\sss+t\mathbf{e}_\jj) \! - \! u_\hh(\sss) \! - \! c_{\hh \jj} \! < \!
u_\jj(\sss) \! - \! u_\hh(\sss) \! - \! c_{\hh \jj} \\
&= u_\ell(\sss)  -  u_\hh(\sss)  -  c_{\hh \ell},
\end{align}
due to $\nabla_{x_\jj}u_\jj(\sss)<0$ by assumption.
Moreover,
\begin{equation}
\begin{aligned}
&\lim_{t\to 0^-} \! \! \! \frac{F_\hh(\sss+t\mathbf{e}_\jj)-F_\hh(\sss)}{t} = \\
&\lim_{t\to 0^-} \! \! \! \frac{[u_\jj\! (\sss+t\mathbf{e}_\jj) \! - \! u_\hh\! (\sss\! ) \! - \! c_{\hh \!  \jj}\! ] \! - \!
[u_\jj\! (\sss\! ) \! - \! u_\hh \! (\sss \! ) \! - \! c_{\hh \!  \jj \! }]}{t} \\
&=\lim_{t\to 0^-} \frac{u_\jj(\sss+t\mathbf{e}_\jj) - u_\jj(\sss)}{t} = \nabla_{x_\jj}u_\jj(\sss) <0,
\end{aligned}
\label{eq:limit_incremental_ratio_b}
\end{equation}
where the first equality holds because for $t<0$ we have
\begin{align}
&u_\jj(\sss+t\mathbf{e}_\jj) \! - \! u_\hh(\sss) \! - \! c_{\hh \jj} \! > \! 
u_\jj(\sss) \! - \!  u_\hh(\sss) \! - \! c_{\hh \jj} \\
&=u_\ell(\sss) \! - \! u_\hh(\sss) \! - \! c_{\hh \ell},
\end{align}
due to $\nabla_{x_\jj}u_\jj(\sss)<0$ by assumption.
From~\eqref{eq:limit_incremental_ratio_a} and~\eqref{eq:limit_incremental_ratio_b} we obtain that
$F_\hh$ is not differentiable at $\sss$,
against what proved in the second part.
Hence we must conclude that there cannot exist $\ell\neq\jj$ such that $\ell\in\A(\hh,\sss)$,
thus $\A(\hh,\sss)=\{\jj\}$.

\ni
4)
Since $F$ is differentiable in $\sss$ by the second part of the proof,
then $\partial F(\sss) = \{ \nabla_\s F(\sss)\}$ is a singleton.
As $\A(\hh,\sss) = \A(\jj,\sss)=\{\jj\}$ by the third part of the proof,
then
\begin{equation}
\begin{aligned}
&u_\jj(\sss)-c_{\hh\jj} > u_\ell(\sss)-c_{\hh\ell}, & &\forall \, \ell \neq \jj, \\
&u_\jj(\sss)-c_{\jj\jj} > u_\ell(\sss)-c_{\jj\ell}, & &\forall \, \ell \neq \jj.
\end{aligned}
\label{eq:singletons}
\end{equation}
As a consequence of~\eqref{eq:singletons} there exists a small enough open ball around $\sss$ where
$F_\jj(\sss)=u_\jj(\sss)-u_\jj(\sss)-c_{\jj\jj}=0$
and $F_\hh(\sss)=u_\jj(\sss)-u_\hh(\sss)-c_{\hh\jj}$.
Thus
\begin{align}
&[\nabla_\s F(\sss)]_{\jj\hh\times \jj\hh}  = \\
&\begin{bmatrix} \frac{\partial F_\jj(\sss)}{\partial \s_\jj} & \frac{\partial F_\jj(\sss)}{\partial \s_\hh} \\
\frac{\partial F_\hh(\sss)}{\partial \s_\jj} & \frac{\partial F_\hh(\sss)}{\partial \s_\hh} \end{bmatrix} =
\begin{bmatrix} 0 & 0 \\ \nabla_{x_\jj}u_\jj(\sss) & -\nabla_{x_\hh}u_\hh(\sss) \end{bmatrix},
\end{align}
whose symmetric part has determinant $0 \cdot \nabla_{x_\hh}u_\hh(\sss)-(\nabla_{x_\jj}u_\jj(\sss))^2/4 < 0$,
which makes
$[\nabla_\s F(\sss)]_{\jj\hh\times \jj\hh}$ indefinite.
Thus $\nabla_\s F(\sss)$ itself is indefinite and $F$ is not monotone in $\S$ due to Proposition~\ref{prop:pos_def_generalized}.
\end{proof}

\section*{\bf Proof of Proposition \ref{prop:proj_converges_to_parallel}}
\begin{proof}
Algorithm~\ref{alg:projection_parallel} is the projection algorithm in~\cite[Alg. 12.1.1]{facchinei2007finite},
applied to VI($\S$,$-u$).
A solution of VI($\S$,$-u$) exists by Proposition~\ref{prop:existence_VI_sol}.
The operator $-u$ is monotone in $\S$, because $\theta$ is concave~\cite[eq. (12)]{scutari2010convex}.
Moreover, due to existence of $\theta$, $L$-Lipschitzianity is equivalent to $(1/L)$-cocoercitivity~\cite[Thm. 18.15]{bauschke:combettes}.
Then, for $\rho<2/L$, Algorithm~\ref{alg:projection_parallel} is guaranteed to converge to
a solution of VI($\S$,$-u$) by~\cite[Thm. 12.1.8]{facchinei2007finite}.
The final claim follows by observing that any Wardrop equilibrium is also an inertial Wardrop equilibrium (Lemma~\ref{lem:parallel_is_inertial}).
\end{proof}

\subsection*{\bf Proof of Theorem \ref{thm:convergence_alg_distr}}
\begin{proof}
First, observe that if $x(0)\in\mc{S}$, then $x(k)$ remains in $\mc{S}$ for all $k\ge 1$. 
This is consequence of the two following observations.
i) At every fixed time-step $k$, and for every pair $i,j$ with $j\in\Nienvy(k)$, the mass $x_{i\rightarrow j}(k)$ is removed from node $i$ and simultaneously added to node $j$ (see Algorithm \ref{alg:better_resp}). Therefore, the total mass must be conserved at each iteration, and so it must be $\sum_{i\in\{1\,\dots,n\}}x_i(k)=\sum_{i\in\{1\,\dots,n\}}x_i(0)=1$.
ii) For every node $i\in\{1,\dots,n\}$, the evolution of $x_i(k)$, as dictated by Algorithm \ref{alg:better_resp}, can be compactly written as
\[
x_i(k+1) = x_i(k) -\sum_{j\in\Nienvy(k)} x_{i\rightarrow j}(k) + \sum_{\ell\,\text{s.t.}\,i\in\mc{E}^{\text{out}}_\ell(k)} x_{\ell\rightarrow i}(k).
\]
Since by assumption $\sum_{j\in\Nienvy(k)} x_{i\rightarrow j}(k) \le x_i(k)$ for every time-step $k$, we have that $x_i(k+1)\ge \sum_{\ell\,\text{s.t.}\,i\in\mc{E}^{\text{out}}_\ell(k)} x_{\ell\rightarrow i}(k)\ge 0$, where the last inequality follows from $x_{\ell\rightarrow i}(k)\ge 0$. Repeating the reasoning for every $k$ ensures that $x_i(k)\ge 0$ at every time-step. Finally, since $\sum_{j\in\Nienvy(k)} x_{i\rightarrow j}(k) \le x_i(k)$, it must be that $x_{\ell\rightarrow i}(k)\le x_{\ell}(k)$. Therefore $\sum_{\ell\,\text{s.t.}\,i\in\mc{E}^{\text{out}}_\ell(k)} x_{\ell\rightarrow i}(k)\le \sum_{\ell\,\text{s.t.}\,i\in\mc{E}^{\text{out}}_\ell(k)} x_{\ell}(k)\le \sum_{\ell\neq i} x_{\ell}(k)$. Hence $x_i(k+1) \le \sum_{\ell \in\{1,\dots,n\}} x_l(k) -\sum_{j\in\Nienvy(k)} x_{i\rightarrow j}(k) \le 1$, where the last inequality follows from the fact that $\sum_{\ell \in\{1,\dots,n\}} x_l(k)=1$ (as shown above) and from the fact that $x_{i\rightarrow j}(k)\ge0$.

We now move our attention to proving the desired convergence statement. To do so, we will show that $\s(k)\to\bs$ such that $\Nienvy(\bs)=\emptyset$ for all $i\in\onen$,
thanks to the equivalence in Proposition~\ref{prop:equiv_out_envy}.
Let us denote for brevity $u_i(k) \defeq u_i(\s_i(k))$ and define
$\mu(k) = \!\!\! \minn{i \in \onen}{\!\! u_i(k)}$.
We show in the following that $\mu(k)$ is a non-decreasing sequence.

First, for any action $i$ we have $\s_i(k+1) - \s_i(k) \le \cmin / \Lmax - \varepsilon$
due to~\eqref{eq:bound_upper}.
Then we can bound the maximum utility decrease
\begin{equation}
\begin{aligned}
&u_i(k+1) - u_i(k) \ge -L |\s_i(k+1) - \s_i(k)| \\ &\ge -L (\cmin / \Lmax - \varepsilon) = -\cmin +L \varepsilon \eqdef -\beta \cmin,
\label{eq:proof_max_utility_decrease}
\end{aligned}
\end{equation}
where the first inequality follows by Lipschitz continuity and we define
$\beta \defeq 1-(L \varepsilon)/\cmin \in ]0,1[$.

Secondly, note that if some action $i$ faces a utility decrease, that is,
if $u_i(k+1) < u_i(k)$,
then it must be $\s_i(k+1) > \s_i(k)$, because $u_i$ is non-increasing.
Then there exists $j$ such that $i \in \Njenvy(\s(k))$.
It follows that
\begin{equation}
\begin{aligned}
&i \text{ faces utility decrease at step } k \Rightarrow \\
&u_i(k) > u_j(k) + c_{ji} \ge \mu(k) + c_\tmin.
\label{eq:proof_i_faces_decrease}
\end{aligned}
\end{equation}
Combining~\eqref{eq:proof_max_utility_decrease} with~\eqref{eq:proof_i_faces_decrease} we obtain
\begin{equation}
\begin{aligned}
&i \text{ faces utility decrease at step } k \Rightarrow \\
&u_i(k+1) > \mu(k) + (1-\beta) c_\tmin,
\end{aligned}
\end{equation}
which implies
$
\mu(k+1) \ge \mu(k).
$
Since $\mu(k)$ is non-decreasing and bounded ($\{u_i\}_{i=1}^n$ are continuous functions in a compact set), there exists a value $\muu$ such that
\begin{equation}
\lim_{k \to \infty} \mu(k) = \muu.
\label{eq:proof_limit}
\end{equation}
We show in the following that there exists an action $\jj$ such that
\begin{equation}
\lim_{k \to \infty} u_\jj(k) = \muu.
\label{eq:proof_limit_uj}
\end{equation}
As $\lim_{k\to \infty} \mu(k) = \muu$, there exists $\hat k$ such that
\begin{equation}
\mu(k) > \muu - c_\tmin(1-\beta)/2, \quad \forall k \ge \hat k.
\label{eq:proof_intermediate}
\end{equation}
Then
\begin{equation}
\begin{aligned}
&i \text{ faces utility decrease at step } k \ge \hat k \Rightarrow \\
&u_i(k) \ge \muu - c_\tmin(1-\beta)/2 + c_\tmin \\
&= \muu + c_\tmin(1+\beta)/2,
\label{eq:proof_max_utility_decrease_after_khat}
\end{aligned}
\end{equation}
where the first inequality follows from combining~\eqref{eq:proof_i_faces_decrease} and~\eqref{eq:proof_intermediate}.
Combining~\eqref{eq:proof_max_utility_decrease} and~\eqref{eq:proof_max_utility_decrease_after_khat} we obtain
\begin{equation}
\begin{aligned}
&i \text{ faces utility decrease at step } k \ge \hat k \Rightarrow \\
& u_i(k+1) \ge \muu - c_\tmin(1-\beta)/2 + c_\tmin(1-\beta) \\ &= \muu + c_\tmin(1-\beta)/2.
\label{eq:proof_possible_utility_after_decrease}
\end{aligned}
\end{equation}
Figure~\ref{fig:proof_convergence} illustrates inequalities~\eqref{eq:proof_max_utility_decrease_after_khat} and~\eqref{eq:proof_possible_utility_after_decrease}.
\begin{figure}[H]
\begin{center}
\includegraphics[width = 0.7\columnwidth]{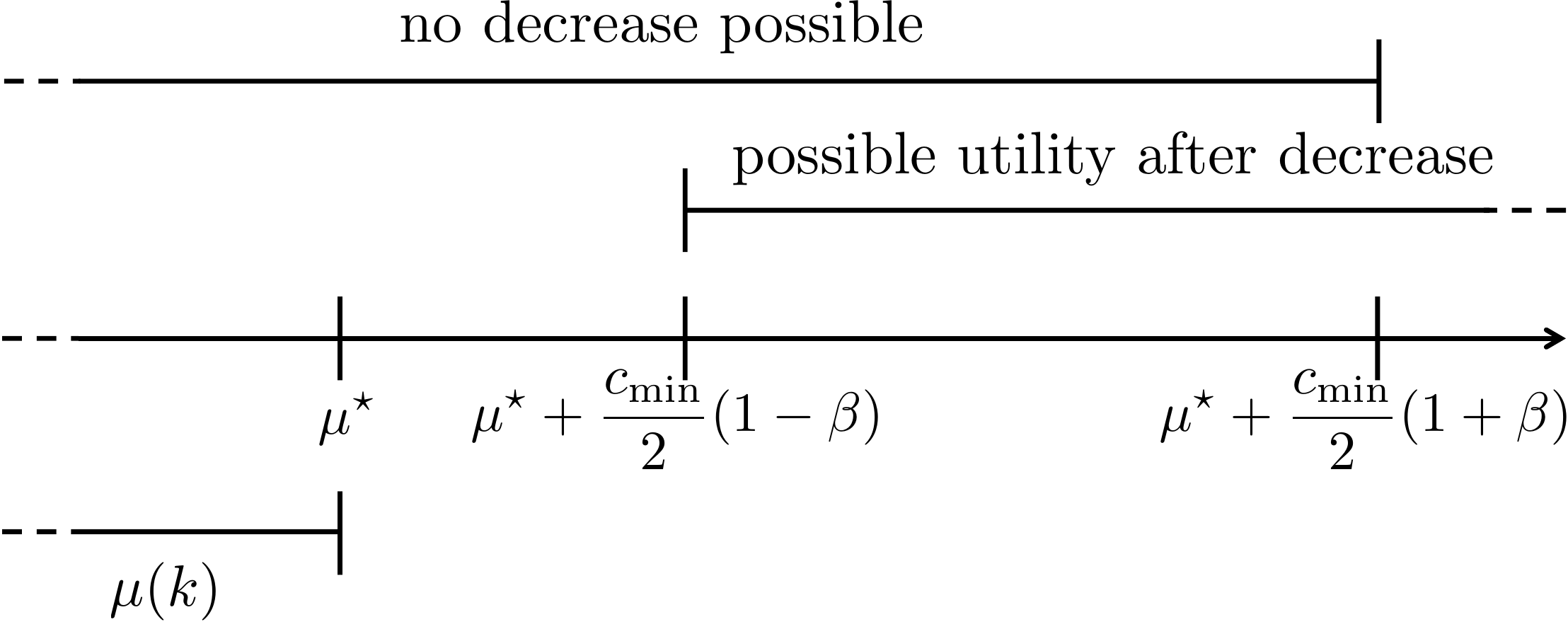}
\end{center}
\caption{Illustration of $\mu(k) \to \muu$ from below and of inequalities~\eqref{eq:proof_max_utility_decrease_after_khat} and~\eqref{eq:proof_possible_utility_after_decrease} after iteration $\hat k$ (with $\beta = 0.5$).}
\label{fig:proof_convergence}
\end{figure}
\noindent
Combining inequalities~\eqref{eq:proof_max_utility_decrease_after_khat} and~\eqref{eq:proof_possible_utility_after_decrease} we obtain that
\begin{equation}
\begin{aligned}
&\exists \, k_1 \ge \hat k \text{ such that } u_i(k_1) \ge \muu + \rho > \muu \Rightarrow \\
&u_i(k) \ge \textup{min} \{\muu + \rho, \muu + c_\tmin(1-\beta)/2 \} \text{ for all } k \ge k_1.
\label{eq:proof_never_go_back}
\end{aligned}
\end{equation}
It then follows
\begin{equation}
\exists \, k_1 \ge \hat k \text{ such that } u_i(k_1) > \muu \Rightarrow \lim_{k \to \infty} u_i(k) \neq \muu.
\label{eq:proof_no_limit_i}
\end{equation}
By~\eqref{eq:proof_no_limit_i} and~\eqref{eq:proof_limit}
it follows that there exists at least an action $\jj$ such that $u_\jj(k) \le \muu \text{ for all } k \ge \hat k$.
Using again~\eqref{eq:proof_limit} and the ``squeeze theorem'' \cite[Thm. 3.3.6]{sohrab2003basic}, we can conclude that $\jj$ satisfies~\eqref{eq:proof_limit_uj}.
Upon defining   
\[\Nienvyrev(\s) = \{i \in \Niin \text{ s.t. } j \in \Nienvy(\s) \},\]
for any $j \in \onen$ and $\s\in\mc{S}$, 
we note that the set $\Nenvyrev_\jj(\s(k))$ 
is empty for $k \ge \hat k$ due to~\eqref{eq:proof_i_faces_decrease} and $u_\jj(k) \le \muu$.
In words, no other action can envy $\jj$ after step $\hat k$.
This implies that $u_\jj(k)$ is a non-decreasing sequence, and in turn $\s_\jj(k)$ is a non-increasing sequence.
As a consequence
\begin{equation}
\lim_{k \to \infty} \s_\jj(k) = \bar \s_\jj \ge 0.
\label{eq:proof_limit_xj}
\end{equation}
If $\bs_\jj=0$, then clearly $\Nenvy_\jj(\bs_\jj,\s_{-\jj})=\emptyset$ by definition, for any $\s_{-\jj}$.
If instead $\bs_\jj>0$,
since $\s_\jj(k+1) \le (1-\tau) \s_\jj(k)$ due to~\eqref{eq:bound_lower},
then convergence is achieved in a finite number of steps.
In other words,
there exists $\tilde k$ such that $\s_\jj(k) = \bar \s_\jj$ for all $k \ge \tilde k$.
In this case, for $k \ge \tilde k$ not only $\Nenvyrev_\jj(\s(k)) = \emptyset$, but also $\Nenvy_\jj(\s(k)) = \emptyset$,
because otherwise $\jj$ would encounter a mass decrease.

Having concluded that there exists $\jj \in \onen$ such that its mass converges (in a finite number of steps if $\bar \s_\jj > 0$),
we propose a last argument to show that there exists $\hh \in \onen \backslash \{ \jj \}$ such that its mass converges to $\bar \s_\hh$
(in a finite number of steps if $\bar \s_\jj, \bar \s_\hh > 0$).
Applying the same argument recursively to $\onen \backslash \{\jj,\hh\}$ concludes the proof.

The last argument distinguishes two cases: $\bar \s_\jj > 0$ and $\bar \s_\jj = 0$.
In the first case $\bar \s_\jj > 0$, we already showed that there exists $\tilde k$ such that
$\Nenvyrev_\jj(\s(k)) = \Nenvy_\jj(\s(k)) = \emptyset$ for all $k>\tilde k$.
Then action $\jj$ has no interaction with any the other action
and considering $k \ge \tilde k$ we apply to $\onen\backslash\jj$
the previous reasoning until equation~\eqref{eq:proof_limit_xj}
to show that there is an action $\hh \in \onen \backslash \{ \jj \}$ with mass that converges to $\bar \s_\hh$
(in a finite number of steps if $\bar \s_\hh > 0$).

In the second case $\bar \s_\jj = 0$. Even though $\Nenvy_\jj$ does not become the empty set at any finite iteration $k$,
the mass $\s_\jj$ becomes so small that transferring mass to the other $n-1$ actions does not have an influence on their convergence.
Proving this requires a cumbersome analysis that does not add much to the intuition already provided.
Let us denote $\eta(k) = \!\!\!\!\! \minn{j \in \onenjj}{\!\!\! u_j(k)}$.
Contrary to $\mu(k)$, the sequence $\eta(k)$ is not non-decreasing in general because the analogous of~\eqref{eq:proof_i_faces_decrease} does not hold, as action $\jj$ could transfer some of its mass to $\onenjj$ thus making their utilities decrease.
Nonetheless, we show that there exists $\etaa$ such that
\begin{equation}
\lim_{k \to \infty} \eta(k) = \etaa.
\label{eq:lim_eta_k}
\end{equation}

To this end, we fix $\epsilon > 0$ and we show that there exists $k^\star$ such that
$| \eta(k) - \etaa| < \epsilon$ for all $k \ge k^\star$.
By definition of $\lim_{k \to \infty} \s_\jj(k) = 0$, there exists $k_\infty$ such that
\begin{equation}
\s_\jj(k) < \epsilon / (2 \Lmax), \quad \forall \, k \ge k_\infty.
\label{eq:limit_xj}
\end{equation}
Let us now construct the sequence
\begin{align}
&\eta^0(k) = \eta(k)+\delta(k), \\
&\delta(k+1) = \delta(k) + \text{max} \{0,\eta(k)-\eta(k+1) \}, \quad \delta(k_\infty) = 0. 
\end{align}
In words, the sequence $\delta(k)$ accumulates the (absolute value of the) decreases of $\eta(k)$ due to $\jj$,
and summing it to $\eta(k)$ results in a sequence $\eta^0(k)$
which is non-decreasing and bounded from above, hence it admits a limit $\etaa$.
By definition,
there exists $k^0$ such that $\eta^0(k) > \etaa-\epsilon/2$ for all $k \ge k^0$.
Moreover, $\delta(k+1) - \delta(k) = \text{max} \{0,\eta(k)-\eta(k+1) \} > 0$ only if $\Nenvy_\jj(\s(k)) \neq \emptyset$ and
in this case $\text{max} \{0,\eta(k)-\eta(k+1) \} \le \Lmax \cdot \sum_{j\neq\jj} \s_{\jj\to j}(k)$.
In words, the only way $\eta(k)$ can decrease is if action $\jj$ transfers some mass to the others, and even then we have a bound on the utility decrease that this can cause.
Summing up
\begin{align}
&\lim_{k \to \infty} \delta(k) = \sum_{k = k_\infty}^\infty \text{max} \{0,\eta(k)-\eta(k+1) \} \\
&\le \Lmax \s_\jj(k_\infty) \underset{\eqref{eq:limit_xj}}{<} \epsilon/2,
\label{eq:delta_bounded}
\end{align}
hence, since $\delta(k)$ is non-decreasing, $\delta(k) < \epsilon/2$ for all $k \ge k_\infty$.
Then for $k \ge \text{max}\{k^\infty,k^0 \}$ it holds
\begin{equation}
\begin{aligned}
\etaa - \eta(k) &= \etaa - \eta^0(k) + \eta^0(k) -\eta(k) \\ &=  \underbracket{\etaa - \eta^0(k)}_{< \epsilon/2} + \underbracket{\delta(k)}_{< \epsilon/2} < \epsilon
\end{aligned}
\end{equation}
which proves~\eqref{eq:lim_eta_k}.

Finally, we want to show that there exists $\hh \in \onenjj$ such that
\begin{equation}
\lim_{k\to \infty} u_\hh(k) = \etaa.
\label{eq:h_attains}
\end{equation}
Consider an action $\ell\neq\jj$ such that
\begin{equation}
\lim_{k \to \infty} u_\ell(\s_\ell(k)) \neq \etaa.
\label{eq:lim_i_neq}
\end{equation}
Since $\eta(k) \to \etaa$,
then $\text{max} \{0,\eta(k)-\eta(k+1) \} \to 0$ as $k\to\infty$.
This, together with $\eta(k)\to\etaa$,
implies that condition~\eqref{eq:lim_i_neq} is equivalent to
the existence of $\theta>0$ such that
for all $k' \ge 0$ there exists $k'' \ge k'$ such that
\begin{equation}
u_\ell(k'') > \etaa + \theta.
\label{eq:larger_theta}
\end{equation}
There are two possibilities in which $\ell$ can face a utility decrease after $k''$,
namely through a mass transfer from some action $\onen \backslash \{\jj,\ell\}$
or through a mass transfer from action $\jj$.
If the mass transfer happens through some action $\onen \backslash \{\jj,\ell\}$,
we can use the same argument of Figure~\ref{fig:proof_convergence} and in particular of implication~\eqref{eq:proof_never_go_back}
to conclude from~\eqref{eq:larger_theta} that
\begin{equation}
u_\ell(k) \ge \textup{min} \{ \etaa + \theta, \etaa + c_\tmin(1-\beta)/2 \}, \; \forall \: k \ge k''.
\label{eq:proof_never_go_back_2}
\end{equation}
If instead the mass transfer happens through $\jj$,
by $\s_\jj(k) \to 0$ one can take $k'$ such that
\begin{equation}
\s_\jj(k) < \theta/(2 \Lmax), \quad \forall k \ge k'
\label{eq:proof_temp_2}
\end{equation}
and take $k''$ such that~\eqref{eq:larger_theta} holds.
Then
\begin{equation}
u_\ell(k) \ge u_\ell(k'') - \Lmax \frac{\theta}{2 \Lmax} > \etaa + \theta - \frac{\theta}{2} = \etaa + \frac{\theta}{2}.
\label{eq:proof_never_go_back_3}
\end{equation}
for all $k \ge k''$,
where the first inequality holds due to Lipschitz continuity and to~\eqref{eq:proof_temp_2},
while the second inequality holds due to~\eqref{eq:larger_theta}.
We can conclude that if~\eqref{eq:lim_i_neq} holds for action $\ell$,
then either~\eqref{eq:proof_never_go_back_2} or~\eqref{eq:proof_never_go_back_3} holds.
Consequently, after $k''$ action $\ell$ does not attain the minimum $\eta(k)$.
If~\eqref{eq:lim_i_neq} holds for all $\ell \in \onen\backslash\jj$, then the minimum $\eta(k)$ is not attained by any action after $k''$, which is a contradiction.
Then there must exist $\hh$ such that~\eqref{eq:h_attains} holds.
With the same argument that led to~\eqref{eq:proof_limit_xj}, we can conclude that there exists $\bs_\hh \ge 0$ such that
$\lim_{k \to \infty} \s_\hh(k) = \bar \s_\hh \ge 0$.
As done for $\jj$, we can conclude that $\Nenvy_\hh=\emptyset$.
\end{proof}

\bibliographystyle{IEEEtran}
\bibliography{bibliography}
\begin{IEEEbiography}[{\includegraphics[width=1.25in,height=1.25in,clip,keepaspectratio]{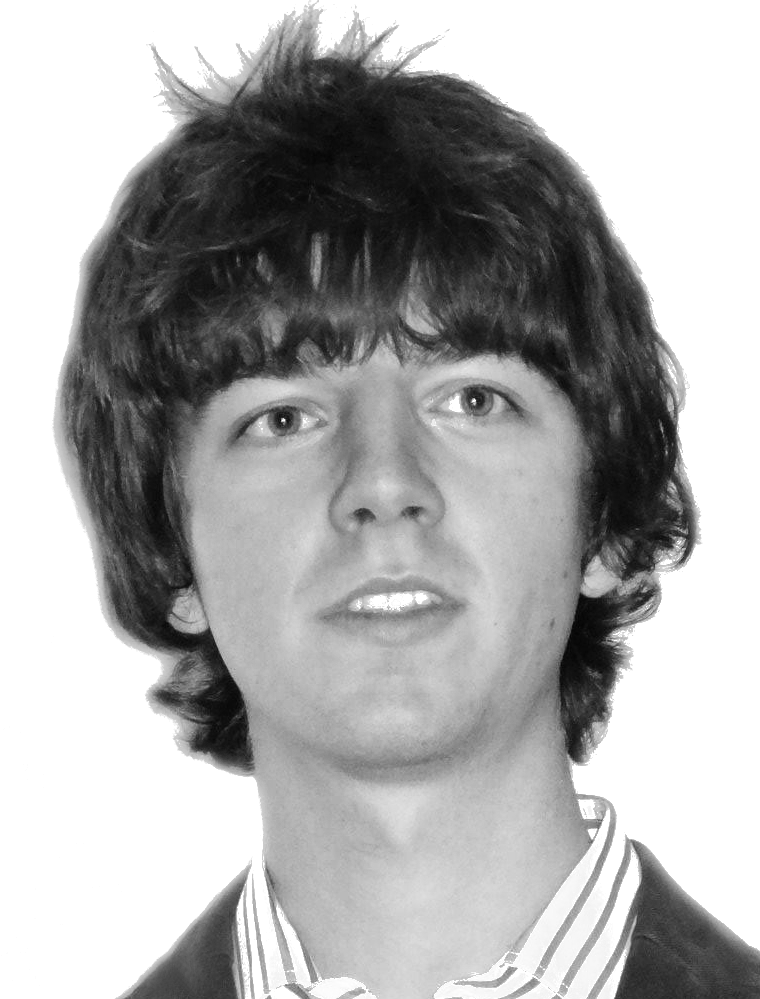}}]
{Basilio Gentile} completed his PhD at the Automatic Control Laboratory at ETH Z\"{u}rich in 2018. He received his Bachelor's degree in Information Engineering and Master's degree in Automation Engineering from the University of Padova, as well as a Master's degree in Mathematical Modeling and Computation from the Technical University of Denmark. In 2013 he spent seven months in the Motion Lab at the University of California Santa Barbara to work at his Master's Thesis. His research focuses on aggregative games and network games with applications to traffic networks and to smart charging of electric vehicles.
\end{IEEEbiography}
\vspace*{-1cm}

\begin{IEEEbiography}[{\includegraphics[width=1.25in,height=1.25in,clip,keepaspectratio]{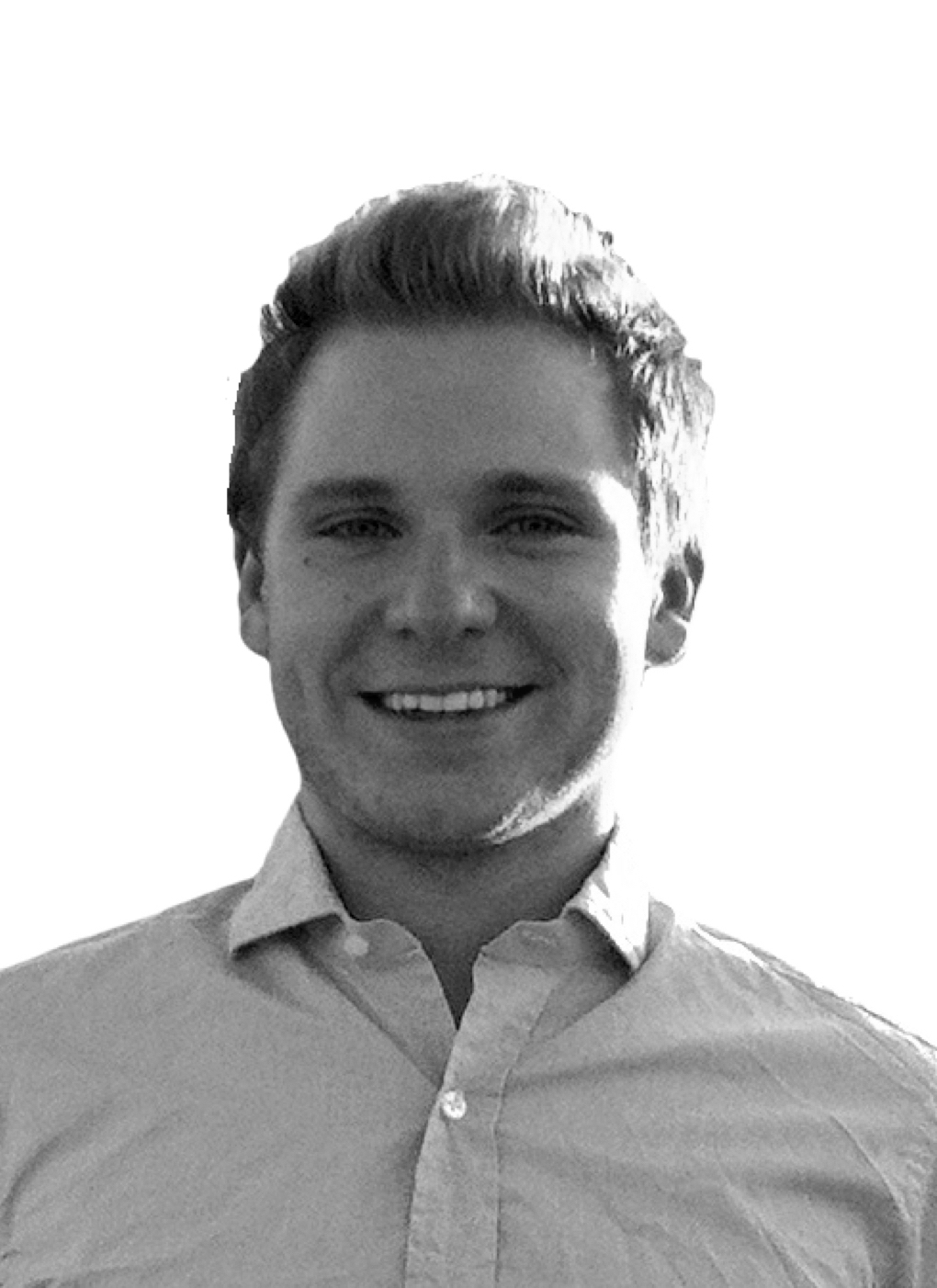}}]{Dario Paccagnan} is a Postdoctoral researcher at the Center for Control, Dynamical Systems, and Computation, U.C. Santa Barbara, USA.
He completed his PhD at the Automatic Control Laboratory, ETH Zurich, in December 2018. 
He received his B.Sc. and M.Sc. in Aerospace Engineering from the University of Padova, Italy, in 2011 and 2014. In the same year he received the M.Sc. in Mathematical Modelling from the Technical University of Denmark, all with Honours. His Master's Thesis was prepared when visiting Imperial College of London, UK, in 2014. From March to August 2017 he has been a visiting scholar at the University of California, Santa Barbara. Dario's research interests are at the interface between distributed control and game theory. Applications include multiagent systems, smart cities and traffic networks.
\end{IEEEbiography}
\vspace*{-1cm}

\begin{IEEEbiography}[{\includegraphics[width=1.25in,height=1.25in,clip,keepaspectratio]{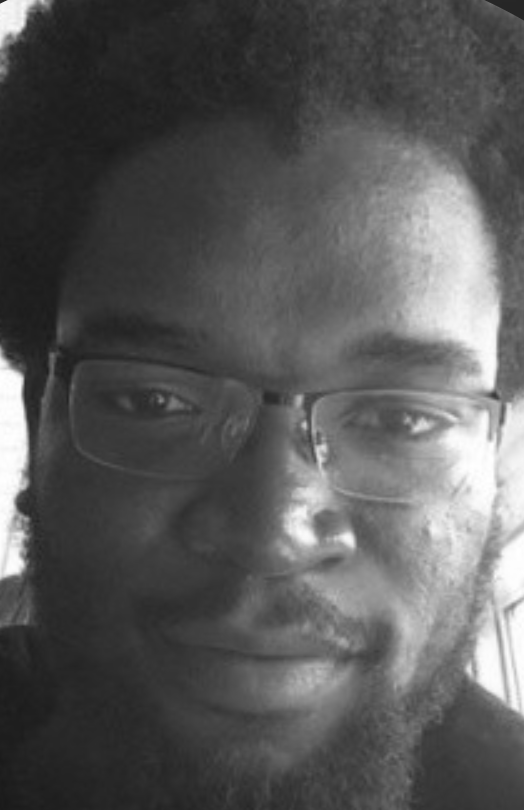}}]{Bolutife Ogunsula} is a software engineer at Bloomberg LP, who completed a master's degree in Robotics, Systems and Control Engineering from ETH Zurich. Prior to that, he got his Bachelor's degree in Electrical and Electronics Engineering form the University of Lagos,
and he worked as a software engineer for Codility.
\end{IEEEbiography}
\vspace*{-1cm}

\begin{IEEEbiography}[{\includegraphics[width=1.25in,height=1.25in,clip,keepaspectratio]{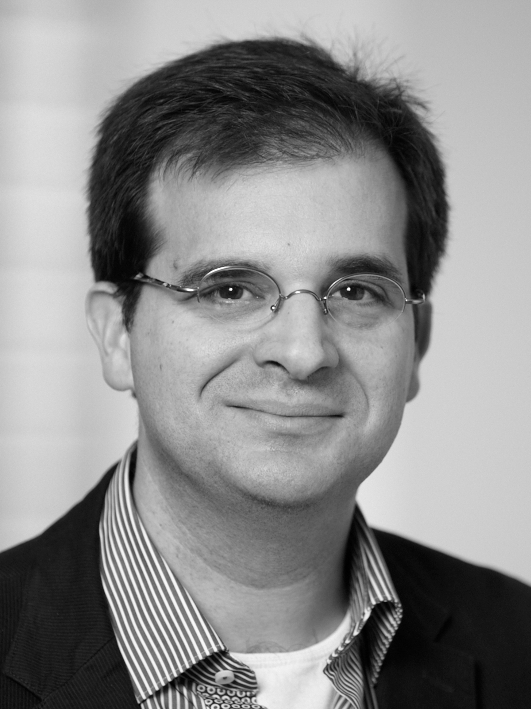}}]{John Lygeros}
completed a B.Eng. degree in electrical engineering in  1990 and an M.Sc. degree in Systems Control in 1991, both at Imperial College of Science Technology and Medicine, London, UK. In 1996 he obtained a Ph.D. degree from the Electrical Engineering and Computer Sciences Department, University of California, Berkeley. During the period 1996-2000 he held a series of research appointments. Between 2000 and 2003 he was a University Lecturer at the Department of Engineering, University of Cambridge, UK. Between 2003 and 2006 he was an Assistant Professor at the Department of Electrical and Computer Engineering, University of Patras, Greece. In July 2006 he joined the Automatic Control Laboratory at ETH Z\"urich, first as an Associate Professor, and since January 2010 as a Full Professor. Since 2009 he is serving as the Head of the Automatic Control Laboratory and since 2015 as the Head of the Department of Information Technology and Electrical Engineering. His research interests include modelling, analysis, and control of hierarchical, hybrid, and stochastic systems, with applications to biochemical networks, automated highway systems, air traffic management, power grids and camera networks. John Lygeros is a Fellow of the IEEE, and a member of the IET and the Technical Chamber of Greece.
\end{IEEEbiography}

\end{document}